\documentclass[usenatbib]{mn2e}
\pdfoutput=1
\usepackage[pdftex]{graphicx}
\usepackage[fleqn]{amsmath}

\defcitealias{1991trcb.book.....D}{RC3}
\defcitealias{2010AJ....139.1566F}{F10}
\defcitealias{2011MNRAS.415.3393F}{F11}
\defcitealias{1988AJ.....96...92A}{AZ88}
\defcitealias{2003MNRAS.340.1317V}{V03}
\defcitealias{2003MNRAS.339..897T}{TMB03+TMK04}

\begin{document}
\title[Extragalactic Globular Cluster Metallicities]{The SLUGGS Survey: Calcium Triplet-based Spectroscopic Metallicities for Over 900 Globular Clusters}
\author[C. Usher et al.]{Christopher~Usher,$^1$\thanks{E-mail: cusher@astro.swin.edu.au} Duncan~A.~Forbes,$^1$ Jean~P.~Brodie,$^2$ Caroline~Foster,$^3$ \newauthor Lee~R.~Spitler,$^1$ Jacob~A.~Arnold,$^2$ Aaron~J.~Romanowsky$^2$, Jay~Strader$^4$\thanks{Menzel Fellow} \newauthor and Vincenzo~Pota$^1$ \\
$^1$Centre for Astrophysics and Supercomputing, Swinburne University of Technology, Hawthorn, VIC 3122, Australia\\
$^2$University of California Observatories, 1156 High Street, Santa Cruz, CA 95064, USA\\
$^3$European Southern Observatory, Alonso de Cordova 3107, Vitacura, Santiago, Chile\\
$^4$Harvard-Smithsonian Centre for Astrophysics, 60 Garden St., Cambridge, MA 02138, USA}
\date{Released 2012 Xxxxx XX}

\pagerange{\pageref{firstpage}--\pageref{lastpage}} \pubyear{2012}
\label{firstpage}
\maketitle

\begin{abstract}
Although the colour distribution of globular clusters in massive galaxies is well known to be bimodal, the spectroscopic metallicity distribution has been measured in only a few galaxies.
After redefining the calcium triplet index--metallicity relation, we use our relation to derive the metallicity of 903 globular clusters in 11 early-type galaxies.
This is the largest sample of spectroscopic globular cluster metallicities yet assembled.
We compare these metallicities with those derived from Lick indices finding good agreement.
In 6 of the 8 galaxies with sufficient numbers of high quality spectra we find bimodality in the spectroscopic metallicity distribution.
Our results imply that most massive early-type galaxies have bimodal metallicity, as well as colour, distributions.
This bimodality suggests that most massive galaxies early-type galaxies experienced two periods of star formation.
\end{abstract}


\section{Introduction}
Optical photometric studies of globular cluster (GC) systems have shown that almost all massive galaxies have bimodal GC colour distributions \citep[e.g.][]{2001AJ....121.2950K, 2001AJ....121.2974L}.
Since extensive spectroscopy has shown the majority of GCs are old ($\geq 10$ Gyr) \citep[e.g.][]{2001ApJ...563L.143F, 2005A&A...439..997P, 2005AJ....130.1315S}, this colour bimodality has usually been interpreted as a metallicity bimodality.
Metallicity bimodality suggests each galaxy has experienced two periods of intense star formation \citep{2006ARA&A..44..193B} which must be accommodated into any scenario of galaxy formation.

However, both \citet{2006BASI...34...83R} and \citet{2006Sci...311.1129Y} showed that a strongly non-linear colour--metallicity relation can produce a bimodal colour distribution from a unimodal metallicity distribution.
If the slope of the colour--metallicity relation flattens at intermediate colours, small differences in metallicity correspond to large differences in colour creating a gap in the colour distribution where none exists in the metallicity distribution.
A unimodal GC metallicity distribution therefore only requires one period of continuous GC formation. 
Thus the true form of the GC metallicity distribution therefore has important ramifications both for globular cluster formation and for galaxy formation.


Although near-infrared (NIR) photometric studies \citep[e.g.][]{2007ApJ...660L.109K, 2008MNRAS.389.1150S, 2012A&A...539A..54C} generally support the interpretation of colour bimodality as metallicity bimodality,  large samples of spectroscopic metallicities are required to settle the question of the form of GC metallicity distributions.
Unfortunately for only a few galaxies have large numbers of GCs been studied spectroscopically.
Although the Milky Way has been long known to host a bimodal metallicity distribution \citep{1979ApJ...231L..19H, 1985ApJ...293..424Z}, the GC metallicity distribution in M31 is less clear.
\citet{2000AJ....119..727B} studied the colour and metallicity distribution of M31 GCs.
They found colour bimodality in $(V - K)$ but not in $(V - I)$.
Using 125 GCs with spectroscopic metallicities they found that the metallicity distribution is bimodal.
They showed that given the large reddening and photometric uncertainties, a unimodal colour distribution would be observed from their spectroscopic metallicity distribution about 75\% of the time.
Using Lick indices \citet{2009A&A...508.1285G} derived the metallicities of 245 M31 GCs; they found that the metallicity distribution is either bimodal or trimodal.
\citet{2011AJ....141...61C} measured the metallicities of 282 GCs in M31 using the strength of iron spectral indices.
Combining their spectroscopic metallicities with 22 GC metallicities obtained using resolved colour magnitude diagrams, they found no statistically significant metallicity bimodality (or trimodality).

\citet{2008MNRAS.386.1443B} measured the metallicity of 207 GCs in NGC 5128, the closest, easily observable giant early-type galaxy, using Lick indices and found a bimodal metallicity distribution.
However the study of \citet{2010ApJ...708.1335W}, which measured the metallicity of 72 GCs in NGC 5128 using Lick indices, did not find statistically significant metallicity bimodality despite finding bimodality in the metallicity sensitive [MgFe]$'$ index and colour.
\citet{2011MNRAS.417.1823A} found evidence of metallicity bimodality in a sample of 112 GCs in the early-type spiral M104 (the Sombrero Galaxy) which they studied spectroscopically. 
\citet{1998ApJ...496..808C} determined the metallicities of 150 GCs in M87 using Lick indices and claimed evidence of bimodality.
From this data set, \citet{2006Sci...311.1129Y} noted that the metallicity sensitive index Mg $b$ had a skewed single peak distribution rather than a bimodal distribution.
Using 47 Lick index measurements by \citet{2003ApJ...592..866C} of the massive elliptical M49's GCs, \citet{2007AJ....133.2015S} found metallicity bimodality.
Although these previous spectroscopic studies suggest GC metallicity bimodality is common in massive galaxies, a larger galaxy sample is required to confirm it.
Due to the large amounts of telescope time required to acquire significant samples of GCs, a more efficient observational technique is required.

The calcium triplet (CaT), at 8498 \AA{}, 8542 \AA{} and 8662 \AA{} in the rest frame, is one of the strongest spectral features in the optical and NIR.
In individual stars the CaT becomes stronger with both increasing metallicity and lower surface gravity \citep{2002MNRAS.329..863C}.
Since giant stars dominate the NIR light of an old population and the surface gravity of metal rich giants is lower than metal poor giants, the strength of the CaT in integrated light increases with metallicity.
The CaT was first used by \citet{1988AJ.....96...92A} to measure the metallicities of Milky Way GCs using integrated light and has been successful at determining the metallicities of individual stars \citep{1997PASP..109..907R, 2008MNRAS.383..183B}.
Single stellar population models \citep[][hereafter V03]{2003MNRAS.340.1317V} show that the CaT index of integrated starlight is sensitive to metallicity while being insensitive to ages greater than 3 Gyr.
Although the CaT index is only weakly sensitive to the IMF for the \citet{1955ApJ...121..161S} IMF and bottom light IMFs such as the \citet{2001MNRAS.322..231K} IMF \citepalias{2003MNRAS.340.1317V}, it is affected by more bottom heavy IMFs such as those found in the most massive elliptical galaxies by \citet{2010Natur.468..940V} and by \citet{2012Natur.484..485C}.

Combining a wide field, red sensitive spectrograph (e.g. \textsc{deimos}, \citealt{2003SPIE.4841.1657F}) and the CaT potentially allows large samples of spectroscopic GC metallicities to be assembled.
Unfortunately the CaT has not yet been proven as a reliable metallicity indicator.
While \citet{1988AJ.....96...92A} found a linear relationship between the strength of the CaT and metallicity over the range of metallicity of Milky Way GCs ($-2 < $[Fe/H]$ < 0$), in the models of \citetalias{2003MNRAS.340.1317V} the CaT saturates at metallicities greater than [Fe/H] $\sim -0.5$.  \citet[hereafter F10]{2010AJ....139.1566F} was the first to use the CaT to determine the metallicity of a large number of extragalactic GCs. 
They used \textsc{deimos} to measure the metallicities of 144 GCs in NGC 1407.
Although they found a bimodal metallicity distribution, the fraction of GCs in the metal poor peak was different than the fraction of GCs in the blue colour peak.
Their measurements of the CaT appeared to be non-linear with colour.
They suggested this conflict is possibly due to saturation of the CaT, a non-linear colour--metallicity relationship or the effects of Paschen lines from hot horizontal branch stars on the CaT measurements.
In addition, they found the brightest blue and red GCs to have similar CaT values.
More recently \citet[hereafter F11]{2011MNRAS.415.3393F} used the same technique to study the metallicites of 57 GCs around NGC 4494.
Unlike in NGC 1407, NGC 4494 showed a linear relation between colour and the CaT index.
Although showing colour bimodality, the CaT strength appeared single peaked in NGC 4494.
A larger sample of CaT measurements in multiple galaxies is thus required to establish whether the CaT can be reliably used to measure metallicity of extragalactic GCs.

\section{Sample}
As part of the ongoing SAGES Legacy Unifying Globulars and Galaxies Survey\footnote{http://sluggs.swin.edu.au} (SLUGGS, Brodie et al. in prep.),  we studied the GC systems of 11 nearby ($D < 30$ Mpc) early-type galaxies. These galaxies cover a range of galaxy masses, morphologies and environments.
Details of the galaxies in this study are given in Table~\ref{tab:galaxies}.
Further details of the observations for NGC 3115 were presented in \citet{2011ApJ...736L..26A}, for NGC 4494 in \citetalias{2011MNRAS.415.3393F} and for the remaining galaxies in \citet{Vince2012}.
We provide summaries of the data acquisition and reduction below. 

\begin{table*}
  \caption{\label{tab:galaxies}Galaxy Properties}
  \begin{tabular}{l c c c c c c c c@{\ }c}
    \hline 
    Galaxy   & Hubble & $V_{sys}$     & $(M - m)$ & $D$    & $M_{K}$  & $A_{B}$ & $N$  & $(g-i)$ &  \\ 
             & Type   & (km s$^{-1}$) & (mag)     & (Mpc)  & (mag)    & (mag)   &      & split   &  \\ 
    (1)      & (2)    & (3)           & (4)       & (5)    & (6)      & (7)     & (8)  & (9)     &  \\ \hline
    NGC 821  & E6     & 1718          & 31.85     & 23.4   & $-24.0$  & 0.48    & 17   & 0.97    &a \\
    NGC 1400 & E0     & 558           & 32.14     & 26.8   & $-23.9$  & 0.28    & 34   & 0.95    &b \\
    NGC 1407 & E0     & 1779          & 32.14     & 26.8   & $-25.3$  & 0.30    & 202  & 0.98    &b \\
    NGC 2768 & S0     & 1353          & 31.69     & 21.8   & $-24.7$  & 0.20    & 49   & 0.95    &b \\
    NGC 3115 & S0     & 633           & 29.87     & 9.4    & $-24.0$  & 0.21    & 122  & 0.93    &c \\
    NGC 3377 & E6     & 690           & 30.19     & 10.9   & $-22.8$  & 0.15    & 84   & 0.86    &b \\
    NGC 4278 & E1     & 620           & 30.97     & 15.6   & $-23.8$  & 0.13    & 150  & 0.96    &b \\
    NGC 4365 & E3     & 1243          & 31.84     & 23.3   & $-25.2$  & 0.09    & 131  & 0.91    &b \\
    NGC 4494 & E2     & 1342          & 31.10     & 16.6   & $-24.1$  & 0.09    & 53   & 0.99    &d \\
    NGC 5846 & E0     & 1712          & 31.92     & 24.2   & $-25.0$  & 0.24    & 54   & 0.96    &b \\
    NGC 7457 & S0     & 844           & 30.55     & 12.9   & $-22.4$  & 0.23    & 7    & ---     &e \\ \hline

  \end{tabular}
	
\medskip
\emph{Notes} Column (1): Galaxy name.
Column (2): Hubble type from the NED database.
Column (3): Systemic velocity from \citet{2011MNRAS.413..813C}.
Column (4) and column (5): distance modulus and distance from \citet{2011MNRAS.413..813C} based on surface brightness fluctuations (SBF) of \citet{2001ApJ...546..681T}. For NGC 4365 the SBF distance of \citet{2007ApJ...655..144M} is used. We assume that NGC 1400 and NGC 1407 are part of the same group and use the mean of their distance moduli.
Column (6): $K$ band absolute magnitude from \citet{2011MNRAS.413..813C} calculated from the 2MASS \citep{2006AJ....131.1163S} total $K$ apparent magnitude and the distance modulus in Column (4), corrected for foreground extinction.
Column(7): $B$ band foreground Galactic extinction from \citet{1998ApJ...500..525S}.
Column (8): Number of GCs with CaT measurements.
Column (9): $(g-i)$ colour split between subpopulations from literature. NGC 4365 shows evidence for colour trimodality; see \citet{2012MNRAS.420...37B}. NGC 7457 appears unimodal; see \citet{2011ApJ...738..113H}. 
Colour split references: a: \citet{2008MNRAS.385..361S}, b: \citet{Vince2012}, c: this work, d: \citetalias{2011MNRAS.415.3393F}, e: \citet{2011ApJ...738..113H}.
NGC 1400, NGC 1407 and NGC 3115 are not part of ATLAS$^{\rmn{3D}}$ sample so their radial velocities were taken from NED while their distances are from \citet{2001ApJ...546..681T} with the same -0.06 mag distance moduli offset.   
Their $K$ band absolute magnitude is also calculated from the 2MASS total $K$ apparent magnitude and the distance in Column (3).
\end{table*}

\subsection{Photometry}
\label{photometry}
\begin{table}
\caption{\label{tab:photo}Photometry}
\begin{tabular}{l l c c} \hline
Galaxy   & Instrument  & Bands & Source \\
(1)      & (2)         & (3)   & (4)    \\ \hline
NGC 821  & Suprime-Cam & $gri$ & \citet{Vince2012} \\
NGC 1400 & Suprime-Cam & $gri$ & \citet{2012MNRAS.423.2177S} \\
NGC 1407 & Suprime-Cam & $gri$ & \citet{2012MNRAS.423.2177S} \\
NGC 2768 & Suprime-Cam & $Riz$ & \citet{Vince2012} \\
         & HST ACS     & $BVI$ & \citet{Vince2012} \\
NGC 3115 & Suprime-Cam & $gri$ & \citet{2011ApJ...736L..26A} \\
NGC 3377 & Suprime-Cam & $gri$ & \citet{Vince2012} \\
NGC 4278 & Suprime-Cam & $BVI$ & \citet{Vince2012} \\
         & HST ACS     & $gz$  & \citet{Vince2012} \\
NGC 4365 & Suprime-Cam & $gri$ & \citet{2012MNRAS.420...37B} \\
         & HST ACS     & $gz$  & \citet{2012MNRAS.420...37B} \\
NGC 4494 & Suprime-Cam & $gri$ & \citetalias{2011MNRAS.415.3393F} \\
NGC 5846 & Suprime-Cam & $gri$ & \citet{Vince2012} \\
         & HST WFPC2   & $VI$  & \citet{1996AJ....112.2448F} \\
NGC 7457 & WIYN        & $BVR$ & \citet{2011ApJ...738..113H} \\
         & Minimosaic  &       & \\  
         & HST WFPC2   & $VI$  & \citet{2008AJ....136..234C} \\ \hline
\end{tabular}

\medskip
\emph{Notes} Column (1): Galaxy name.
Column (2): Telescope and instrument.
Column (3): Photometric bands.
Column (4): Reference to source of photometry.
\end{table}

For most of the galaxies in our sample we used Subaru Suprime-Cam \citep{2002PASJ...54..833M} $gri$ imaging, supplemented with other wide field imaging and HST imaging.
The sources of our photometry is given in Table~\ref{tab:photo}.
We note that the $g$ band imaging in NGC 821 and NGC 5846 and $B$ band imaging from NGC 4278 were observed in poor seeing.
All photometry is corrected for the effects of Galactic extinction using the extinction maps of \citet{1998ApJ...500..525S}. 
For galaxies and GCs that lack $gri$ imaging we used empirical colour conversions to give all GCs equivalent $(g-i)$ colours.
These conversions are given in Appendix~\ref{colours}.
Unlike our spectroscopy, the photometry used in this paper is heterogeneous - it was taken from different studies with different zero points and aperture corrections. 
From the wide-field, ground-based imaging we selected GC candidates using cuts in colour--colour space.
From HST imaging we selected GC candidates using cuts in colour-size space.
We refer to the references for each photometric study for details of the selection.
Colour magnitude diagrams and colour histograms of the photometric candidates are shown in Figure~\ref{fig:mag}.

\subsection{Spectroscopy}

Between 2006 and 2012, we used \textsc{deimos} on the Keck II telescope in multi-object mode to obtain spectra of large numbers of candidate GCs.
These observations were primarily designed to obtain large numbers of GC radial velocities rather than to study GC stellar populations.
Early observations used a central wavelength of 7500 \AA{} while later observations used a central wavelength of 7800 \AA{}.
All observations used the 1200 line mm$^{-1}$ grating and 1 arcsec slits.
This set up yields a resolution of $\Delta \lambda \sim 1.5$ \AA{} and covers the CaT region for almost all slits.
We observed 2 to 11 slit masks per galaxy with typical exposure times of 2 hours per mask.

The \textsc{deimos} data were reduced using the \textsc{idl} \textsc{spec2d} pipeline \citep{Newman2012, 2012ascl.soft03003C}.
The pipeline performs flat fielding using internal flats, wavelength calibration using ArKrNeXe arcs and local sky subtraction.
In addition to the science spectrum, a fully propagated error array and a background sky spectrum are also produced by the pipeline for each object.
After running the pipeline, heliocentric velocity corrections were applied.
Radial velocities were measured using the \textsc{IRAF}\footnote{\textsc{IRAF} is distributed by the National Optical Astronomy Observatory, which is operated by the Association of Universities for Research in Astronomy (AURA) under cooperative agreement with the National Science Foundation.} procedure \textsc{fxcor}.
To separate GCs from stars \citet{Vince2012} used the friendless algorithm of \citet{2003MNRAS.346L..62M} for their galaxies.
For NGC 3115 we used a velocity cut of 325 km$^{-1}$ to seperate stars and GCs while in the case of NGC 4494 the GC system is well separated in velocity space from stars.
We measured the signal-to-noise ratio (S/N) per \AA ngstrom of the spectra using the mean S/N per pixel in the range 8400 to 8500 \AA.

To increase the number of GCs with CaT measurements that have literature metallicities, we supplemented our \textsc{deimos} data with long slit spectra of two M31 GCs obtained using \textsc{lris} on Keck I \citep{1995PASP..107..375O} as poor weather targets on 2010 August 14.
These observations used the 600/10000 grating, a central wavelength of 8448 \AA{} and a 1.5 arcsec slit.
B012-G064 was observed for $3 \times 300$s while B225-G280 was observed for $4 \times 300$s.
These \textsc{lris} observations were reduced in the standard manner using \textsc{IRAF}. 

\begin{figure*}
	\begin{center}
		\includegraphics[width=504pt]{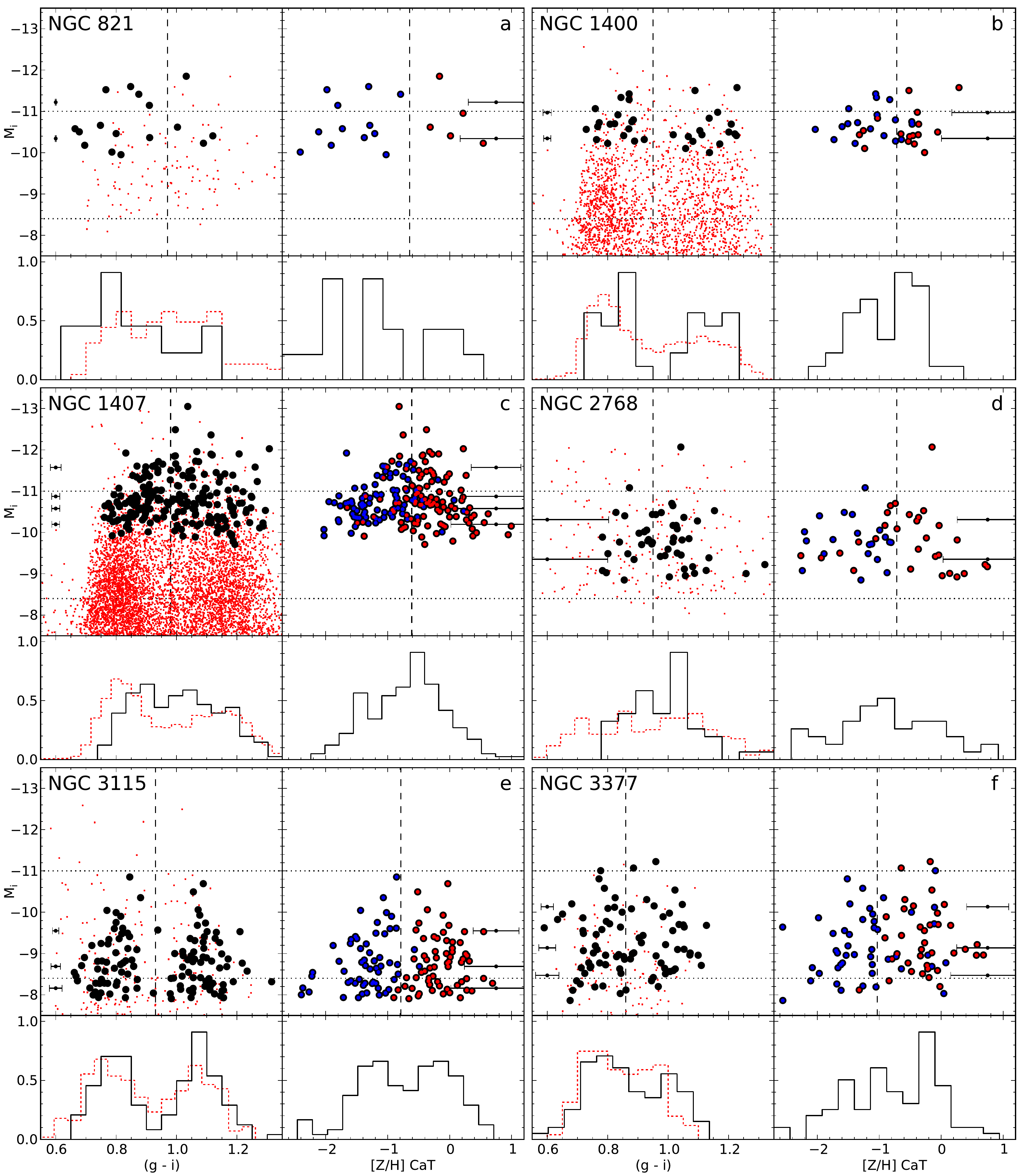}
		\caption{\label{fig:mag} Colour magnitude and metallicity magnitude diagrams for each galaxy.
For each galaxy the upper left plot is a colour magnitude diagram, the upper right a CaT metallicity magnitude diagram, the lower left a normalised colour histogram and the lower right a normalised metallicity histogram.
Black filled circles and solid lines are GCs with CaT metallicities; red small points and dotted lines are GC candidates.
In the metallicity magnitude diagrams the centres of the points are colour coded blue or red depending on the photometric colour split.
Blue GCs are metal poor; red GCs are metal rich.
Median colour and metallicity error bars are plotted to the left and right.
The horizontal dashed lines show $M_{i} = -8.4$, the turnover magnitude from \citet{2010ApJ...717..603V}, and $M_{i} = -11$, comparable to $\omega$ Cen's absolute magnitude ($M_{i} = -11.25$, \citealt{1996AJ....112.1487H}) which corresponds to a mass of $\sim 1 \times 10^{6}$ M$_{\sun}$.
On the colour magnitude diagrams the vertical dashed line is the colour split between red and blue subpopulations from Table~\ref{tab:galaxies}.
On the metallicity magnitude diagrams the vertical dashed line is colour split transformed into metallicity using Equation~\ref{eq:colourmetal}.
Note that the metallicities of blue GCs remain on the metal poor side of colour division and vice versa.}

	\end{center}
\end{figure*}
\addtocounter{figure}{-1}
\begin{figure*}
	\begin{center}
		\includegraphics[width=504pt]{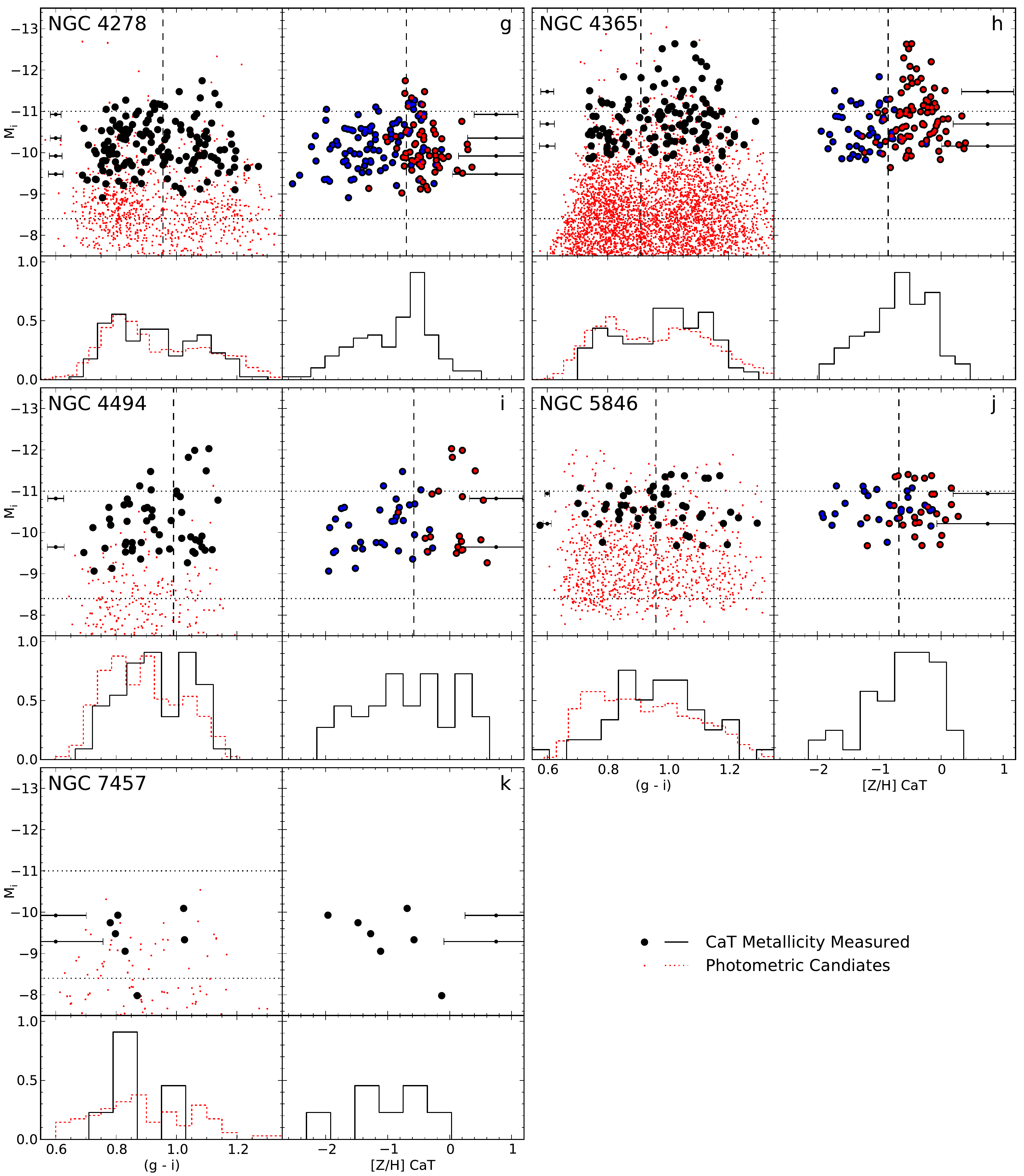}
		\caption{\emph{Continued}}

	\end{center}
\end{figure*}

\subsection{Comparison sample}
\label{sec:comparison}
To compare our CaT metallicities with previous work we assembled a sample of literature metallicities based on Lick indices \citep{1994ApJS...94..687W}.
We used the metallicity studies of \citet{2007AJ....134..391C} for 17 GCs in NGC 1407, \citet{2002A&A...395..761K} and Norris (priv. comm.) for 14 GCs NGC 3115, \citet{2005AJ....129.2643B} for 21 GCs in NGC 4365 and \citet{2008AJ....136..234C} for 13 GCs in NGC 7457.
All these studies derived metallicities in the same manner using $\chi^{2}$ minimization of Lick indices with the single stellar population (SSP) models of \citet{2003MNRAS.339..897T} and \citet{2004MNRAS.351L..19T} (both hereafter \citetalias{2003MNRAS.339..897T}).
We removed object 25a from \citet{2002A&A...395..761K} as its colour is much bluer than the other GCs and its radial velocity is low enough to be a galactic star.
For these four galaxies we used our own GC photometry.

For M31 we used the metallicities from \citet{2011AJ....141...61C} who used an empirical relationship to derive the metallicities from the strength of Fe Lick indices.
Since the \citet{2011AJ....141...61C} metallicities are on the \citet{1997A&AS..121...95C} scale, Equation~\ref{eq:cg97metal} was used to convert the metallicities to the same \citetalias{2003MNRAS.339..897T} scale used elsewhere in this paper.
We used the M31 SSDS GC photometry of \citet{2010MNRAS.402..803P}. 
Due to the variable foreground and internal reddening of M31 we only used the 25 GCs with literature extinction values derived from resolved colour magnitude diagrams.
We used reddening values from \citet{2005AJ....129.2670R}, \citet{2006ApJ...650L.107G}, \citet{2007ApJ...655L..85M}, \citet{2009A&A...507.1375P}, \citet{2011AJ....141...61C}, and \citet{2011A&A...531A.155P}.
For GCs in common with multiple studies we used the average value; however values from other studies were preferred to values from \citet{2011AJ....141...61C}.
\citet{2005AJ....129.2670R} provides two sets of reddenings: the values adopted and the values fitted as a free parameter.
Except for B045-G108 and B311-G033 we used the adopted values; for these two clusters the fitted values provide better agreement with the colour--metallicity trend seen in other clusters.
To account for the large uncertainties associated with the reddening correction we have added a 0.05 dex in quadrature to the photometric errors on the colours.
In summary we have Lick index-based metallicities for 90 GCs in five galaxies, of which we have measured the CaT strength of 31 GCs.

\section{Calcium Triplet Measurement}
The CaT lies in a region of strong sky lines.
Although the \textsc{deimos} pipeline does a good job subtracting the sky, the sky line residuals prevent line indices from being accurately measured directly on the observed spectra.
To remove the sky line residuals we used the technique of \citetalias{2010AJ....139.1566F} and fit stellar templates to the spectra.
The stellar templates, which cover a range of metallicities and a range of spectral classes (11 giants and 2 dwarfs ranging from F to early M), were observed using \textsc{deimos} from 2007 November 12th to 14th using an instrument set up comparable to the one used for the GCs.
The templates were fitted using the \textsc{pPXF} pixel fitting code of \citet{2004PASP..116..138C} which also fits radial velocity, velocity dispersion and continuum shape.
We fitted the 8425 to 8850 \AA{} rest frame spectral range and masked areas with strong sky lines from the fit.

Before measuring the strength of the CaT on the fitted spectra, the continuum was normalised.
Unlike in \citetalias{2010AJ....139.1566F} the continuum was normalised without human intervention.
Instead, the continuum was modelled using a linear combination of Chebyshev polynomials of order zero to eight.
Regions of the fitted spectra around known spectral lines were masked.
A continuum model was then fit using least squares.
Pixels more than 0.4 \% below or 1 \% above the model were rejected and the model recomputed.
This process was repeated until no further pixels were rejected.
After the initial fit the pixels were only masked if they deviated from the model.
The same index definition (8490.0\AA{} to 8506.0\AA{}, 8532.0\AA{} to 8552.0\AA{} and 8653.0\AA{} to 8671.0\AA{}) as \citetalias{2010AJ....139.1566F} was used to measure the strength of the CaT on the normalised spectra.

\label{caterrors}
In \citetalias{2010AJ....139.1566F} the errors in the measured calcium index were estimated using the S/N of the spectrum.
Although the S/N of the spectra dominates the errors, both metallicity and radial velocity do have an effect on the errors.
At high metallicities stronger weak metal lines make it harder to fit the continuum while at certain radial velocities such as 200 km s$^{-1}$ CaT lines fall on sky lines. 
In this work the uncertainties of the index measurements were derived by using a Monte Carlo resampling technique.
One hundred realisations of the spectra were created from the fitted spectra and the observed error array.
The spectral fitting, continuum normalisation and index measurement process was repeated for each resampling.
The 68th percentiles of the resamplings were used as confidence intervals.
This procedure gives asymmetric error bars.
After measuring the strength of the CaT and calculating the associated uncertainty, each spectrum was inspected visually to check the results of the automated fitting and normalisation.
Spectra with large velocity dispersions, $\sigma > 100$ km s$^{-1}$, were rejected as likely being contaminated by background galaxy light.
Spectra with poor sky subtraction where also rejected.
The software used to measure the CaT index and calculate errors is available from the authors upon request.

\section{Analysis}
\subsection{Metallicity scales}
Studies of extragalactic GC metallicities have used different metallicity scales.
Since they include the effects of varying $\alpha$ element enhancement, the SSP models of \citetalias{2003MNRAS.339..897T} have been commonly used in extragalactic GC studies to derive metallicities from Lick indices.
However studies that have used empirical relations based on the properties of Milky Way GCs tend to use the metallicity values from the \citet{1996AJ....112.1487H} catalogue.
Although originally based on the \citet{1984ApJS...55...45Z} metallicity scale, the Harris catalogue has evolved with improved abundances and is now based \citep{harrisMWcatalogue} on the \citet{2009A&A...508..695C} metallicity scale which uses large numbers ($\sim 2000$) of high resolution spectra of individual GC stars.
A challenged faced both by SSP models and empirical relations is that the Milky Way lacks the metal rich GCs seen in massive early-type galaxies.

Since \citetalias{2003MNRAS.339..897T} models have been most commonly used in extragalactic globular stellar population studies, we used their metallicity scale in this paper.
To place metallicity measurements and models that used different metallicity scales on the same scale as the one defined by the \citetalias{2003MNRAS.339..897T} models we derived empirical relations using Milky Way GCs.
\citet{2007MNRAS.379.1618M} measured stellar population parameters of Milky Way GCs using \citetalias{2003MNRAS.339..897T} models.
We used their total metallicities ([Z/H]) and the iron abundances ([Fe/H]$_{\text{C09}}$) given by \citet{2009A&A...508..695C} to derive a linear relationship between the \citetalias{2003MNRAS.339..897T} metallicity scale and the \citet{2009A&A...508..695C} iron abundance scale:
\begin{equation}\label{eq:carrettametal}
\text{[Z/H]} = 0.829 \text{[Fe/H]}_{\text{C09}} -0.083 \text \, .
\end{equation}
This relation has a rms difference of 0.138 dex.
We combined the previous relation and the relation provided by \citet{2009A&A...508..695C} between their iron abundance scale and the \citet{1997A&AS..121...95C} iron abundance ([Fe/H]$_{\text{CG97}}$) scale, to give a relation between the \citet{1997A&AS..121...95C} iron abundance scale and the \citetalias{2003MNRAS.339..897T} scale:
\begin{equation}\label{eq:cg97metal}
\text{[Z/H]} = 0.943 \text{[Fe/H]}_{\text{CG97}} - 0.085 \, . 
\end{equation}
This relation has a rms difference of 0.151 dex.
We also derived a relation between the \citetalias{2003MNRAS.339..897T} metallicity scale and the scale defined by the 2003 version of the Harris catalogue. Using the \citet{2007MNRAS.379.1618M} measurements we found the following relation:  
\begin{equation}\label{eq:harrismetal}
\text{[Z/H]} = 0.939 \text{[Z/H]}_{\text{H03}} + 0.043 \, .
\end{equation}
This relation has a rms difference of 0.135 dex.

\begin{figure}
\begin{center}
\includegraphics[width=240pt]{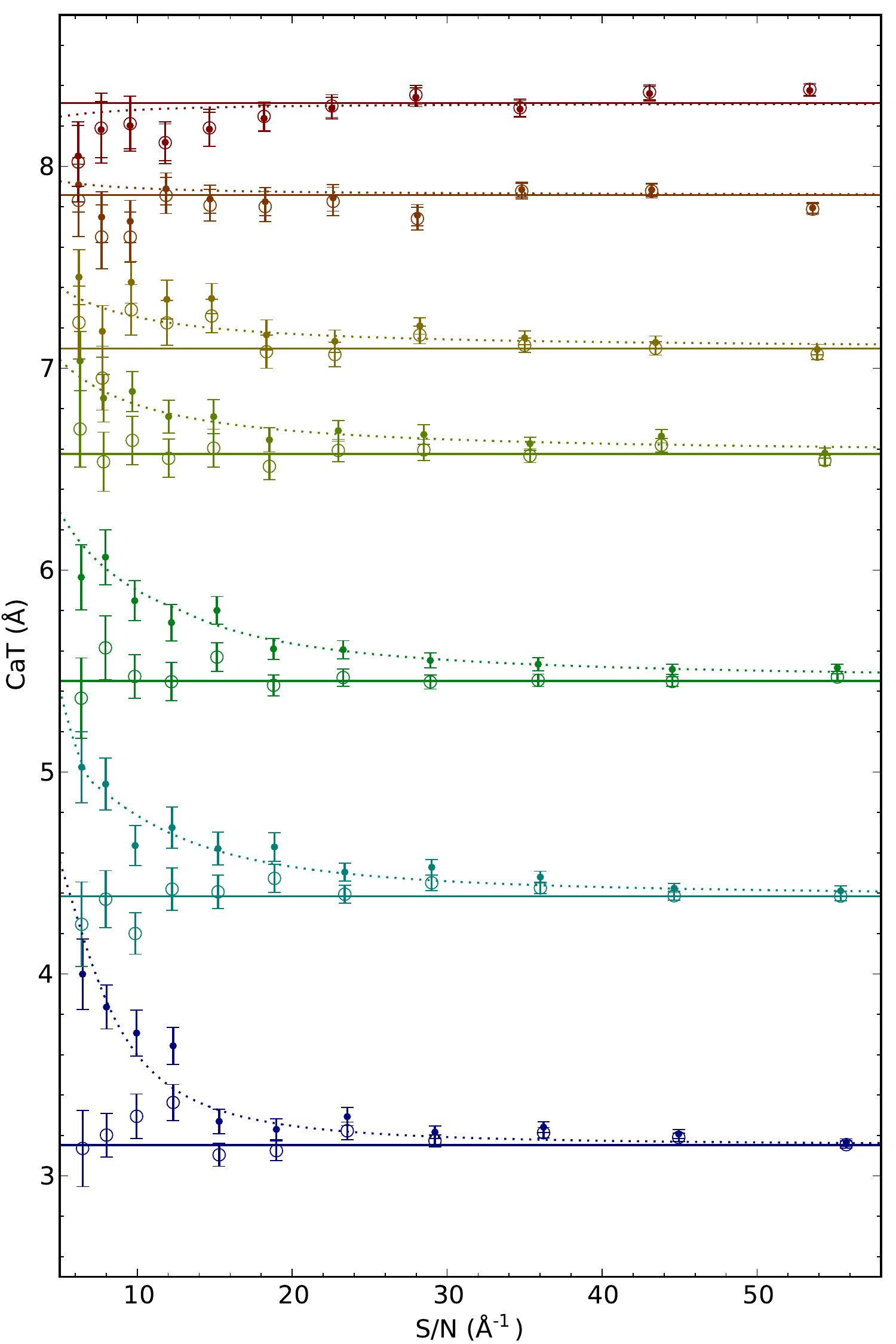}
\caption{\label{fig:catcorrection} Correction for the S/N bias in the CaT index values.
The vertical axis is the CaT index value, the horizontal axis is the measured S/N.
The small points are the mean measured CaT values from spectra generated from \citet{2003MNRAS.340.1317V} model spectra at different metallicities and S/N ratios.
The solid lines are the CaT values measured at high S/N for each model metallicity.
The dotted lines are Equation~\ref{eq:correction}, a fit to difference between the measured points and the high S/N values.
The hollow circles show the measured values after using Equation~\ref{eq:correction} to correct them.
The colours in this plot scale with the metallicities of the model spectra used with blue being metal poor and red being metal rich.
The S/N bais appears to be caused by \textsc{pPXF}.}
\end{center}
\end{figure}

\subsection{Biases}
\label{bias}
Before deriving metallicities we checked whether our CaT measurements were biased by velocity dispersion, radial velocity or S/N.
We used \citetalias{2003MNRAS.340.1317V} SSP model spectra, with an age of 12.6 Gyr and a \citet{2001MNRAS.322..231K} IMF, to create spectra with arbitrary velocity dispersion, radial velocity and S/N at each of the model metallicities.
The \citetalias{2003MNRAS.340.1317V} models were used because they well match the resolution of the \textsc{deimos} spectra ($\sim 1.5$ \AA{} for both).
The \citetalias{2003MNRAS.340.1317V} models use the isochrones of \citet{2000A&AS..141..371G} and the empirical stellar library of \citet{2001MNRAS.326..959C}.
With increasing velocity dispersion a bias towards lower index values was seen for the highest metallicities.
However, this effect only became important for velocity dispersions larger than those seen in GCs.
A bias towards higher index values was seen at certain radial velocities such as 200 km s$^{-1}$.
This is likely caused by the CaT lines falling on masked skyline regions.  
A strong bias to higher index values was seen at S/N $< 20$ \AA$^{-1}$ with the effect being more severe for metal poor spectra.
The bias appears to reverse for the most metal rich spectra.
The difference is seen in the fitted spectra so it is due to the pPXF fitting rather than the normalisation process. 
To correct for this effect the mean CaT value was measured on spectra generated from the model spectra for a range of S/N ratios and metallicities.
By using the mean index values at large ($\sim 300$ \AA$^{-1}$) S/N as the true index values, the following relation was used to correct for the S/N bias:

\begin{multline}\label{eq:correction}
\text{CaT} = \text{CaT}_{text{raw}} + a \times (S/N)^{b} \\
a = \left\{\begin{array}{c c}
-22.5\, \text{CaT}_{text{raw}} + 122  &\text{CaT}_{text{raw}} < 4.96 \\
-3.33\, \text{CaT}_{text{raw}} + 27.0 &\text{CaT}_{text{raw}} > 4.96 \end{array}\right. \\
b = \left\{\begin{array}{c c}
-0.343\, \text{CaT}_{text{raw}} + 3.20  &\text{CaT}_{text{raw}} < 5.80 \\
0.031\, \text{CaT}_{text{raw}} + 1.03 &\text{CaT}_{text{raw}} > 5.80 \end{array} \right. \, .
\end{multline}

The S/N bias and the applied correction can be seen in Figure~\ref{fig:catcorrection}. 
All further CaT values in this study are corrected for the S/N bias.
We restrict our analysis to spectra with a S/N greater than 8 \AA$^{-1}$.
Using a S/N cut of 10 or 12 \AA$^{-1}$ in our analysis gave identical results so we chose to use the lower S/N cut to increase our sample sizes.
Colour magnitude diagrams and colour histograms of the GCs with CaT measurements are shown in Figure~\ref{fig:mag}.

\begin{figure}
	\begin{center}
		\includegraphics[width=240pt]{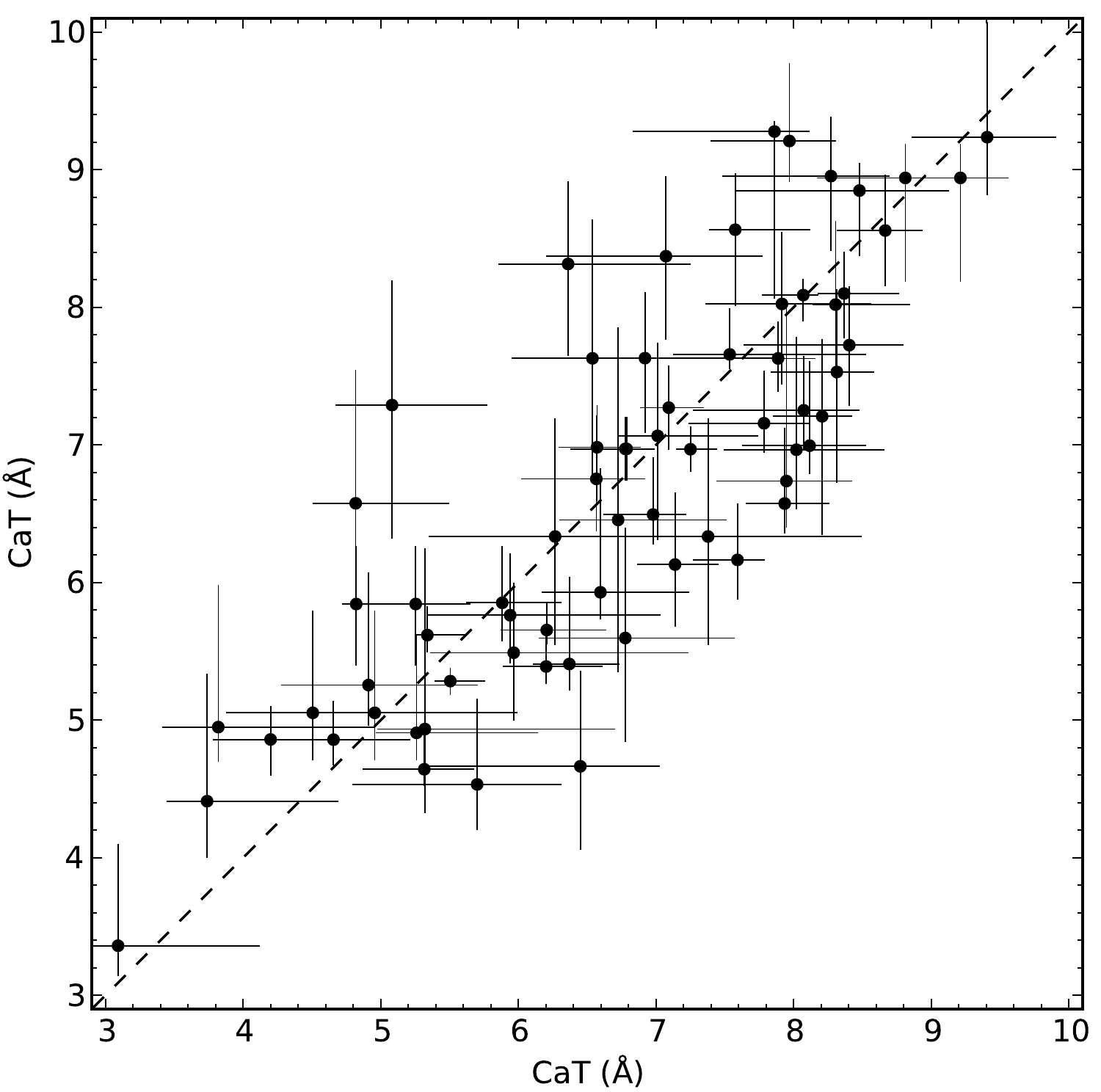}
		\caption{\label{fig:repeats} Repeated measurements of CaT index values.
The dashed line is one-to-one.
The rms difference of the repeated measurements is 0.84 \AA.
The error bars include the 0.255 \AA{} systematic error.}
	\end{center}
\end{figure}

\subsection{Repeated Measurements}
To assess the quality of the CaT index measurements we compare the 68 repeated measurements in Figure~\ref{fig:repeats}.
The reduced $\chi^{2}$ value of the repeated measurements is 1.41 suggesting that that Monte Carlo resampling process is underestimating the errors.
Adding a systematic error of 0.255 \AA{} in quadrature lowers the reduced $\chi^{2}$ to 1.00
We included this systematic error in future analysis.
For repeated objects the simple mean of the CaT indices was used for further analysis.

\begin{figure}
	\begin{center}
   
		\includegraphics[width=240pt]{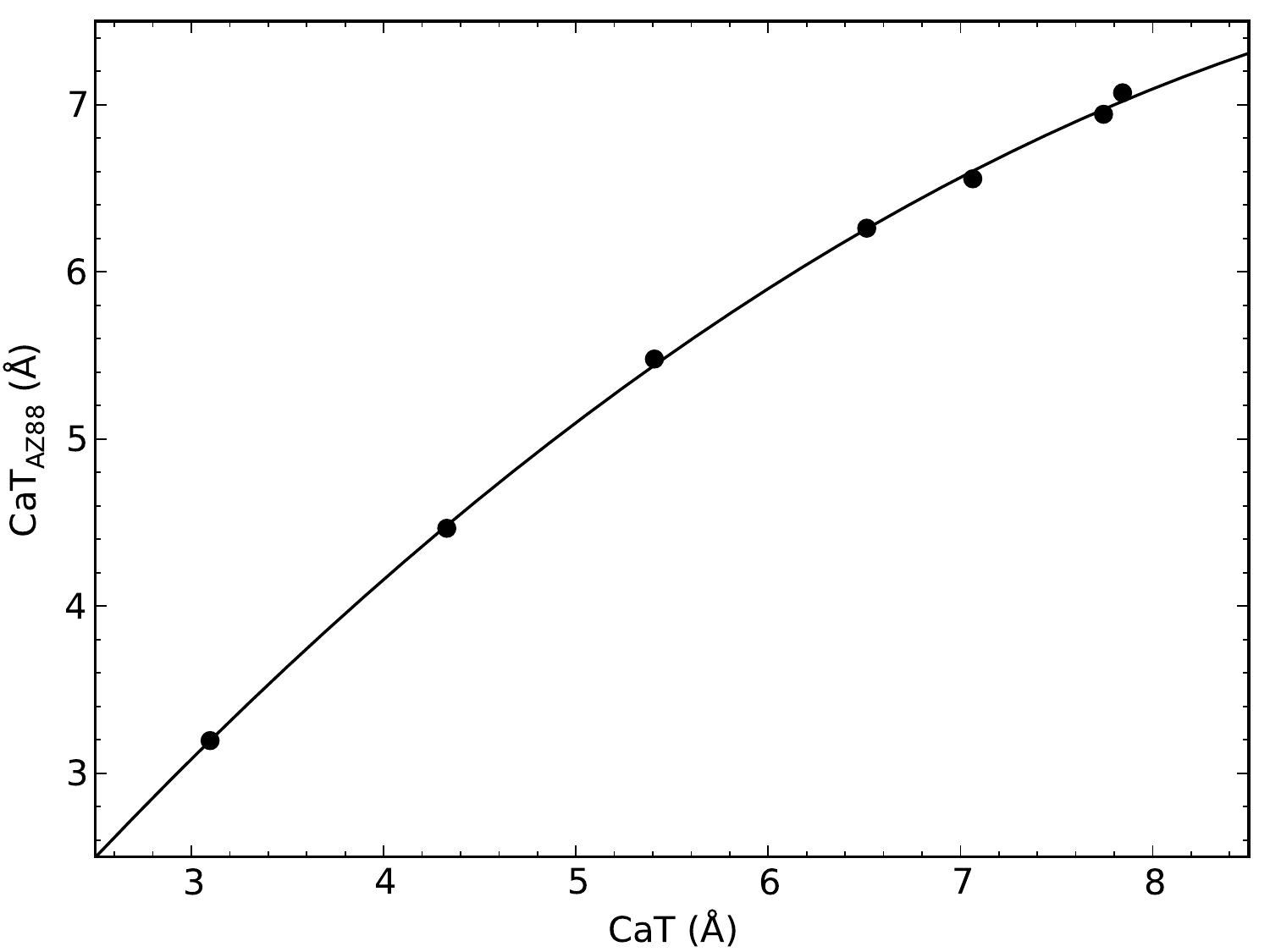}
		\caption{\label{fig:usheranzcal}Relation between the CaT index of \citet{1988AJ.....96...92A} and our index definition. Both indices are measured on spectra from the 12.6 Gyr, \citet{2001MNRAS.322..231K} IMF, single stellar population models of \citet{2003MNRAS.340.1317V}. The points are the measured values and the solid line is a quadratic fit (Equation~\ref{eq:usheranzcal}).}

	\end{center}
\end{figure} 
\begin{figure}
	\begin{center}

		\includegraphics[width=240pt]{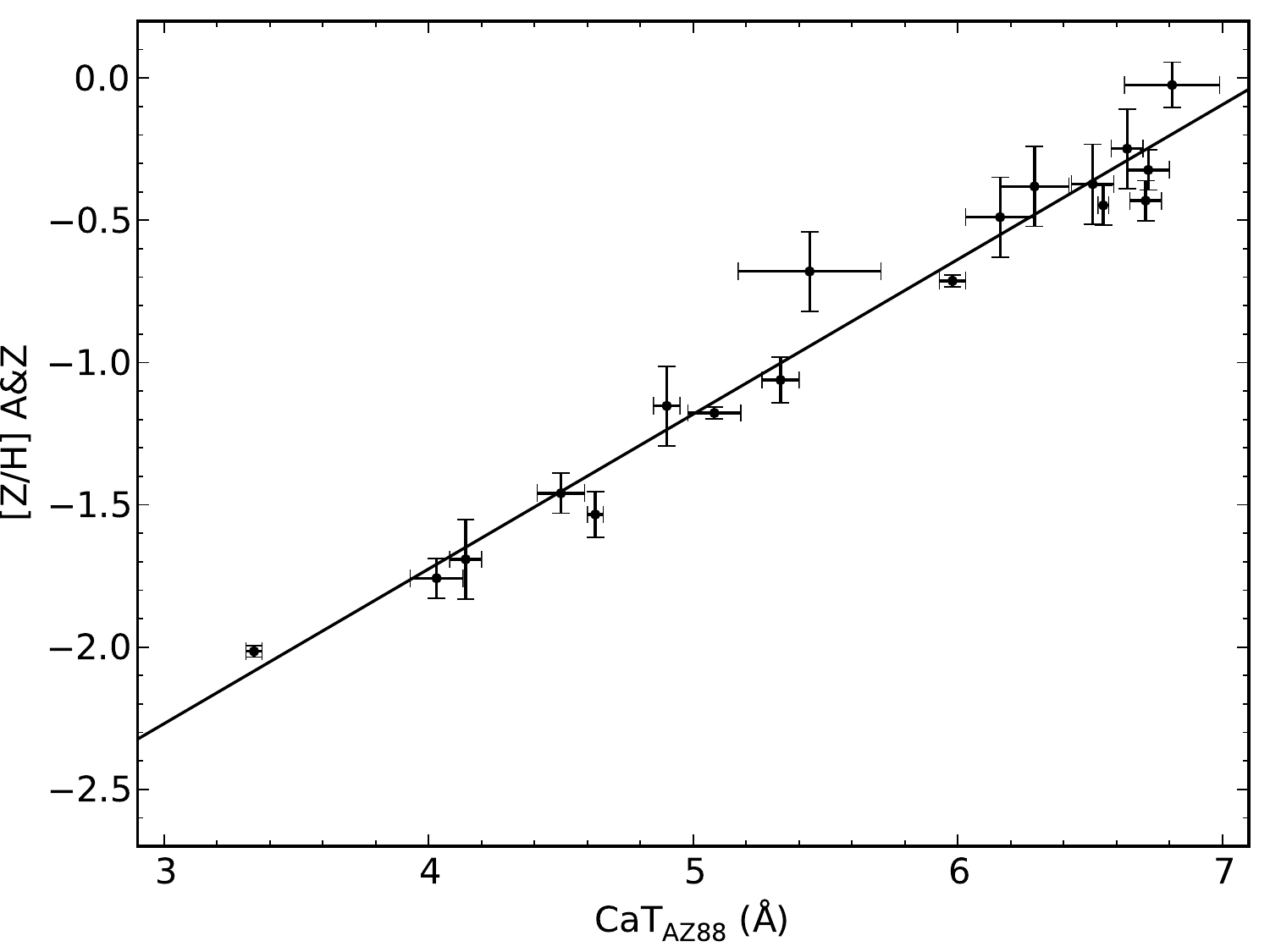}
		\caption{\label{fig:catanzfecal}Relation between metallicity ([Z/H]) and the CaT index definition of \citet{1988AJ.....96...92A} for 18 Milky Way GCs. The [Z/H] values are from \citet{2009A&A...508..695C} converted to the same metallicity scale used in the models of \citet{2003MNRAS.339..897T} using Equation~\ref{eq:carrettametal}. The solid line is a linear fit (Equation~\ref{eq:catanzfecal}) which has a rms scatter of 0.11 dex.}
	\end{center}
\end{figure}

\subsection{Calibration}
\label{calib}
\begin{figure}
	\begin{center}

		\includegraphics[width=240pt]{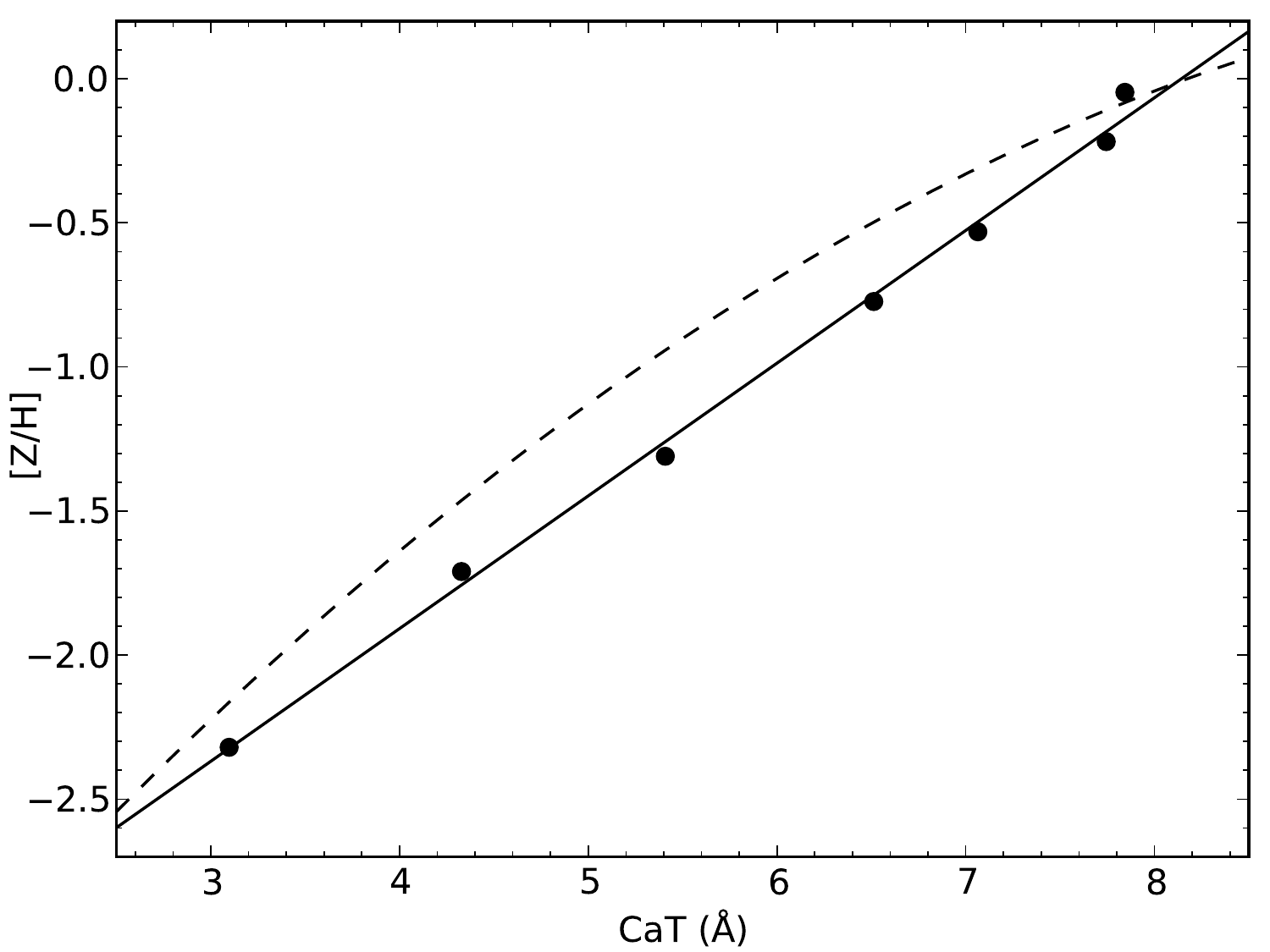}
		\caption{\label{fig:catsspcal}Relation between metallicity ([Z/H]) and our CaT index.
The points are the index values measured on the 12.6 Gyr, \citet{2001MNRAS.322..231K} IMF, single stellar population models of \citet{2003MNRAS.340.1317V}.
The model metallicities are corrected for the local stellar $\alpha$ element enhancement pattern using Equation~\ref{eq:vazdekismetal}.
The solid line is a linear fit to these points (Equation~\ref{eq:catfecal}). For comparison the dashed line is the relation derived using the measurements of \citet{1988AJ.....96...92A} (Equation~\ref{eq:usheranzcal}).
Although both calibration techniques agree at high and low metallicities, there is a $\sim 0.3$ dex difference in [Z/H] at CaT $=5.5$ \AA.}
	\end{center}
\end{figure}

Two approaches were taken to calibrate the relation between the CaT index and metallicity.
First we used the same approach as \citetalias{2010AJ....139.1566F} using the CaT index strengths of \citet[hereafter AZ88]{1988AJ.....96...92A} to perform an empirical calibration.
We derived a relation between our CaT index and the \citetalias{1988AJ.....96...92A} index definition (CaT$_{\text{AZ88}}$).
We measured both our index and the \citetalias{1988AJ.....96...92A} index on \citetalias{2003MNRAS.340.1317V} SSP models with an age of 12.6 Gyr and a \citet{2001MNRAS.322..231K} IMF.
The following quadratic relation was fit to the measurements and is shown in Figure~\ref{fig:usheranzcal}:
\begin{multline}\label{eq:usheranzcal}
\text{CaT}_{\text{AZ88}} = (-0.068 \pm 0.007) \text{CaT}^{2} + (1.55 \pm 0.08) \text{CaT} \\ + (-0.96 \pm 0.22) \, .
\end{multline}
We also rederived the relationship between metallicity and the \citetalias{1988AJ.....96...92A} index.
Unlike \citetalias{2010AJ....139.1566F} who used the 2003 edition of the \citet{1996AJ....112.1487H} catalogue for metallicities, we used the metallicities provided in Appendix A of \citet{2009A&A...508..695C} converted to the \citetalias{2003MNRAS.339..897T} scale together with the CaT$_{\text{AZ88}}$ values provided in Table 6 of \citetalias{1988AJ.....96...92A}.
Like \citetalias{2010AJ....139.1566F} we ignore highly reddened clusters.
Fitting a straight line gives:
\begin{equation}\label{eq:catanzfecal}
\text{[Z/H]} = (0.54 \pm 0.02) \text{CaT}_{AZ88} + (-3.90 \pm 0.14) \, .
\end{equation}
This relation is plotted in Figure~\ref{fig:catanzfecal} and has a rms scatter of 0.11 dex.
Combining Equations \ref{eq:usheranzcal} and \ref{eq:catanzfecal} gives:
\begin{multline}\label{eq:cal}
\text{[Z/H]} = (-0.037 \pm 0.004) \text{CaT}^{2} + (0.84 \pm 0.06) \text{CaT} \\ + (-4.42 \pm 0.20)
\end{multline}
which is plotted in Figure~\ref{fig:catsspcal}. This Figure gives the relations between our CaT index and metallicity for both of our calibration approaches.

Our second approach to calibrate the relation between metallicity and CaT index was to use the \citetalias{2003MNRAS.340.1317V} SSP.
We measured our index values on the model spectra with an age of 12.6 Gyr and a \citet{2001MNRAS.322..231K} IMF and fitted a linear relation between the metallicity of the models and the measured index value giving a relation of:
\begin{equation}\label{eq:catfecal}
\text{[Z/H]} = (0.461 \pm 0.013) \text{CaT} + (-3.750 \pm 0.080)
\end{equation}
which is also plotted in Figure~\ref{fig:catsspcal}.
The rms of this relation is 0.048 dex.
Unlike \citetalias{2003MNRAS.339..897T}, \citetalias{2003MNRAS.340.1317V} does not correct for the effects of varying [$\alpha$/Fe] in the local stellar neighbourhood.
Following \citet{2007MNRAS.379.1618M} we correct the model metallicities for this effect using:
\begin{equation}\label{eq:vazdekismetal}
\text{[Z/H]} = \left\{\begin{array}{cc@{\ }c@{\ }c@{\ }c@{\ }c}
\text{[Z/H]}_{\text{V03}} &&&\text{[Z/H]}_{\text{V03}}&<&-1.0 \\
&&&& \\
\dfrac{\text{[Z/H]}_{\text{V03}} - 1.28}{1.28} &-1.0&<&\text{[Z/H]}_{\text{V03}} &<&0.28 \\
&&&& \\
\text{[Z/H]}_{\text{V03}} - 0.28 & 0.28&<&\text{[Z/H]}_{\text{V03}}&& \end{array} \right. \, .
\end{equation}

\begin{figure*}
\begin{center}
\includegraphics[width=504pt]{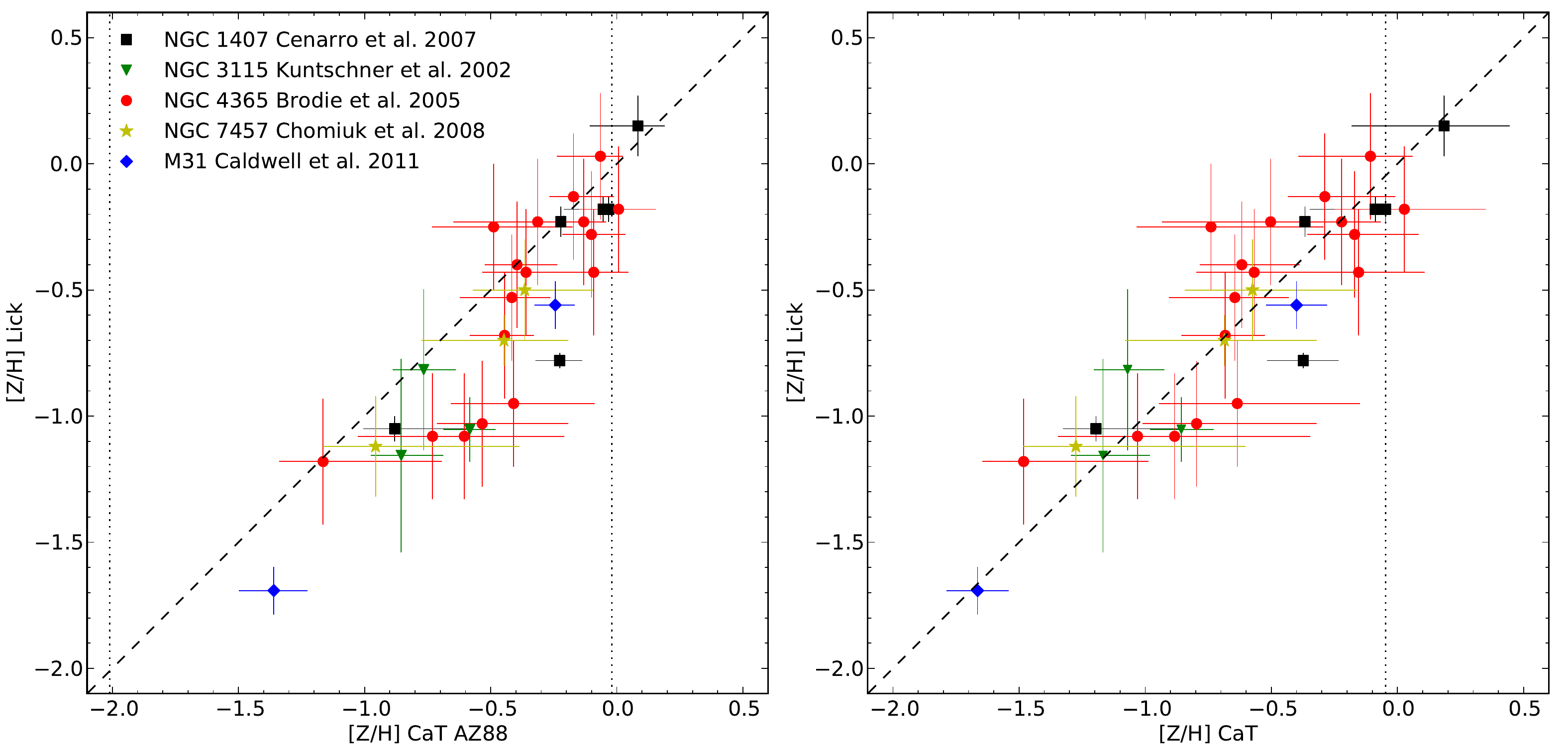}
\caption{\label{fig:compareLickCaT} \emph{Left} CaT metallicities derived using \citet{1988AJ.....96...92A} based calibration compared to literature Lick index-based metallicities.
Different galaxies and studies are denoted by different colours and symbols.
The dashed line is one-to-one and the vertical dotted lines correspond to the highest ([Z/H]$ = -0.02$) and lowest ([Z/H]$ = -2.02$)  metallicity Galactic GCs used to derive the relation between index strength and metallicity.
For metallicities of [Z/H]$ < -0.5$, the metallicities given by empirical calibration are about 0.3 dex higher than the literature values.
The rms scatter is 0.27 dex and the reduced $\chi^{2}$ value is 2.26.
\emph{Right} CaT metallicities derived using single stellar population (SSP) model of \citet{2003MNRAS.340.1317V} compared to literature metallicities.
Different galaxies and studies are denoted by different colours and symbols.
The dashed line is one-to-one while the vertical dotted line corresponds to the highest metallicity SSP model ([Z/H]$ = -0.05$) used to derive the relation between index strength and metallicity.
The lowest metallicity SSP population model used ([Z/H]$ = -2.32$) is too low to appear on this plot.
The rms scatter is 0.20 dex and the reduced $\chi^{2}$ value is 0.59.} 
\end{center}
\end{figure*}

\label{catlick}
To check our results and to choose the better calibration we compared the metallicities we derived with literature spectroscopic measurements.
We restricted our analysis to GCs with a S/N greater than 8 \AA$^{-1}$ and literature ages greater than 7 Gyr.
The comparison of the CaT metallicities derived using the \citetalias{1988AJ.....96...92A} empirical calibration with literature metallicities is given in Figure~\ref{fig:compareLickCaT}. 
For most metallicities ([Z/H] $< -0.5$) this calibration gives metallicities 0.3 dex higher than literature values.
However at higher metallicities there is good agreement.

The comparison of the CaT metallicities derived using the \citetalias{2003MNRAS.340.1317V} SSP models with literature metallicities is also given in Figure~\ref{fig:compareLickCaT}.
Unlike the empirical calibration, the SSP-based calibration approach gives metallicities in good agreement with literature values across the whole range of observed metallicities.
Since the reduced $\chi^{2}$ value is significantly lower for the SSP calibration (0.59) than for the empirical calibration (2.26) and gives better qualitative agreement, we choose to use the SSP calibration approach for the rest of the analysis (except when comparing to the previous measurements of \citetalias{2010AJ....139.1566F} and \citetalias{2011MNRAS.415.3393F} in Section~\ref{previous}).
Metallicity magnitude diagrams and metallicity histograms for each galaxy are plotted in Figure~\ref{fig:mag}.
Colours, magnitudes, CaT index measurements and CaT metallicities for all measured GCs are given in Table~\ref{tab:data} which is available in its entirety online.

\begin{figure}
	\begin{center}

		\includegraphics[width=240pt]{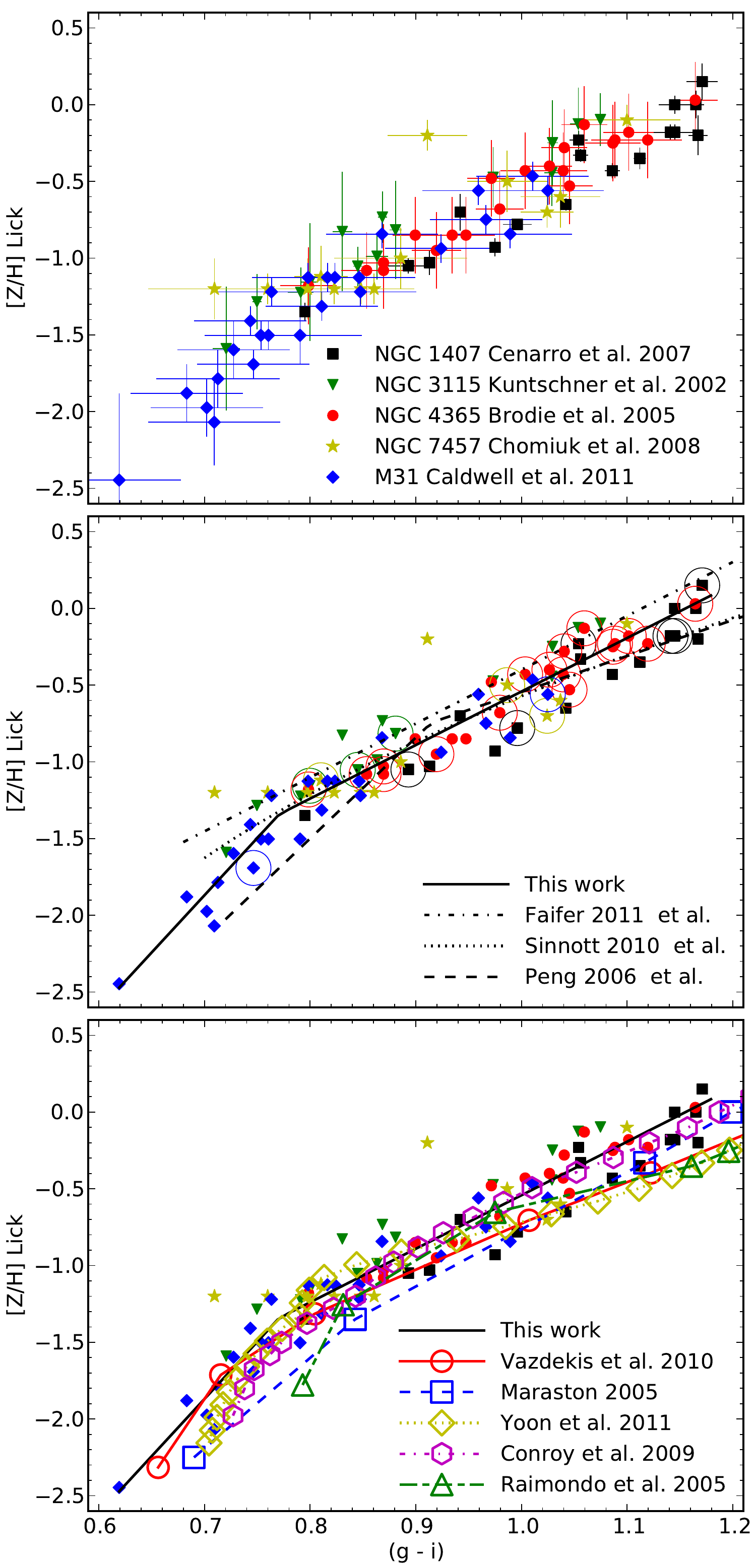}
		\caption{\label{fig:compareLickColour}Globular cluster colour--metallicity relations.
Note that different studies used different techniques to derive metallicities and colours (see text for details) although we attempt place them on the same scales.
\emph{Top} Different colours and symbols refer to different galaxy studies.
For [Z/H] $> -1.4$ the colour appears to be linear while for [Z/H] $< -1.4$ the relation steepens.
\emph{Middle} Empirical colour--metallicity relations.
The solid line is our broken linear fit to the data while the other lines are the colour--metallicity relations of \citet{2011MNRAS.416..155F}, \citet{2010AJ....140.2101S} and \citet{2006ApJ...639...95P}.
However, the different relations agree within the uncertainties of each relation above $(g - i) \sim 0.8$.
We have measured CaT metallicities for the circled points.
\emph{Bottom} Single stellar population (SSP) model colour--metallicity relations.
The solid line is our broken linear fit to the data.
The coloured lines are the SSP models of \citet{2010MNRAS.404.1639V}, \citet{2005MNRAS.362..799M}, \citet{2011ApJ...743..150Y}, \citet{2009ApJ...699..486C} and \citet{2005AJ....130.2625R}.
All models are $\sim 13$ Gyr old; see text for details of the other stellar population parameters used.}

	\end{center}
\end{figure}

\subsection{Colour--metallicity relation}
\label{gettingcolourmetal}
To compare the colours of GCs with their CaT metallicities we constructed an empirical colour--metallicity relation using literature metallicities (Section~\ref{sec:comparison}) and our own photometry.
We fitted the following broken linear function (See Figure~\ref{fig:compareLickColour}): 
\begin{equation}\label{eq:colourmetal}
\text{[Z/H]} = \left\{\begin{array}{r} (7.46 \pm 1.28) \times (g - i) + (-7.09 \pm 1.01) \\
(g - i) < 0.77 \\
\\
(3.49 \pm 0.12) \times (g - i) + (-4.03 \pm 0.11) \\
(g - i) > 0.77 \end{array}\right.
\end{equation}
The RMS of this colour--metallicity relation is 0.17 dex.
Versions of this relation in other optical colours are available in Appendix~\ref{othercolourmetals}.
Combining colour--metallicity data from multiple galaxies may introduce scatter due to variations in the photometric zero points and reddening corrections. 
We note that the break in the colour--metallicity relation occurs at roughly the same metallicity ([Z/H] $\sim -1.4$) as the break in the metallicity--iron index relation used by \citet{2011AJ....141...61C}.

\begin{figure*}
	\begin{center}
		\includegraphics[width=504pt]{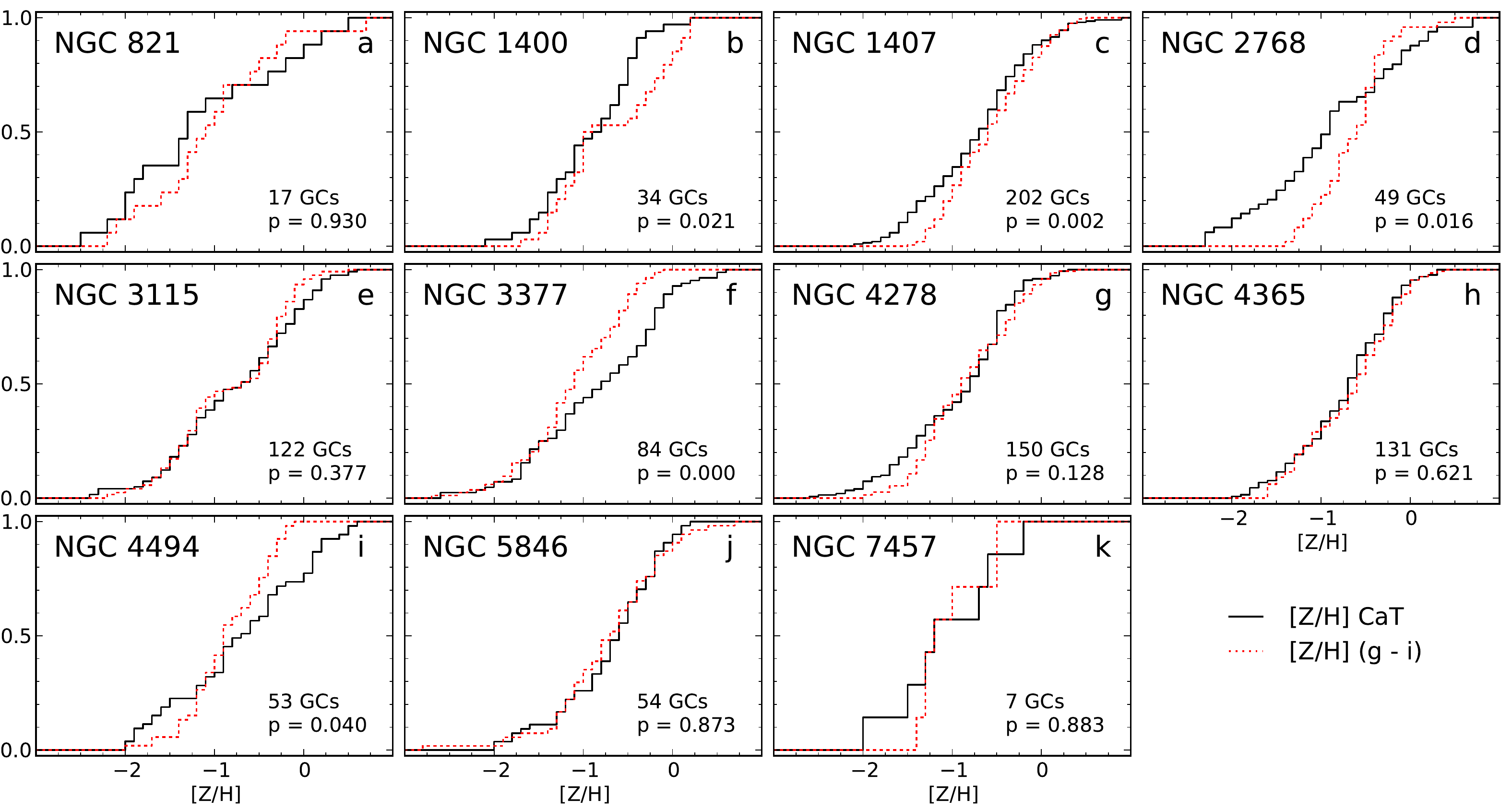}
		\caption{\label{fig:zhist} Cumulative metallicity histograms for each galaxy.
The solid black lines are the metallicities derived using the CaT; the dashed red lines are the same GCs with metallicities derived using $(g - i)$ colours.
The probability given by the KS test that the two samples are drawn from the same distribution is given in the lower right corner of each plot along with the number of GCs with CaT measurements.}

	\end{center}
\end{figure*}

A comparison of the metallicity distribution derived from the CaT compared to the distribution derived from the colours is given by the cumulative histograms in Figure~\ref{fig:zhist}.
To quantify this comparison we ran the Kolmogorov-Smirnov (KS) test on the CaT metallicities and the colour-based metallicities for the same objects.
A non parametric test, the KS test measures the probability that two samples are drawn from the same distribution.
We take a probability of less than $p = 0.05$ as evidence that the CaT-based and colour-based metallicities are drawn from different distributions and inconsistent with one another. 
The results of the KS tests are given in Figure~\ref{fig:zhist}.

\subsection{Testing bimodality}
\label{GMM}
To test for colour and metallicity bimodality we used the Gaussian Mixture Modelling (GMM) code of \citet{2010ApJ...718.1266M}.
An improved version of the widely used KMM code of \citet{1994AJ....108.2348A}, GMM calculates both the best fitting unimodal and best fitting bimodal distributions to the data.
It then runs a parametric bootstrap to determine whether the bimodal distribution is preferred.
\citet{2010ApJ...718.1266M} discuss detecting bimodality and note that looking for an improved fit by two Gaussians over one is not so much a test of bimodality as a test of non-Gaussianity.
Therefore further requirements must be met to declare a population bimodal.
\citet{2010ApJ...718.1266M} suggest using the relative separation of the fitted peaks, $D$, and the kurtosis of the sample to decide whether a sample is bimodal.
They require $D > 2$ and negative kurtosis in addition to a significant bootstrap test result for a population to be bimodal. 
We adopt the same requirements. 
In line with previous bimodality studies we consider a distribution bimodal if bimodality is preferred over unimodality with a probability of $p \geq 0.9$.

In addition to running GMM on the colours and metallicities (both colour-based and CaT-based) of the GCs with calcium triplet measurements, we ran GMM on the colours of the GC photometric candidates.
The photometric candidates were restricted to those in the absolute magnitude range $-13.5 < M_{i} < -8.4$, the fainter limit corresponding to the turnover magnitude \citep{2010ApJ...717..603V} which is the peak of the GC magnitude distribution.
We also limited the galactocentric radial range to radii less than that of the most distant spectroscopically confirmed GC.
In the case of NGC 4494 we followed \citetalias{2011MNRAS.415.3393F} and excluded candidates closer than 0.5 arcmin to the galaxy.
We did not make any corrections for contamination; such corrections would be different in each galaxy.
We only ran GMM when the sample size was greater than or equal to 40.
Results of running GMM on the metallicity distributions are given in Table~\ref{tab:gmm-metal} and on the colours in Table~\ref{tab:gmm-colour}.
The GMM metallicity results are plotted in Figure~\ref{fig:metalmetal}.

\begin{figure*}
	\begin{center}
		\includegraphics[width=504pt]{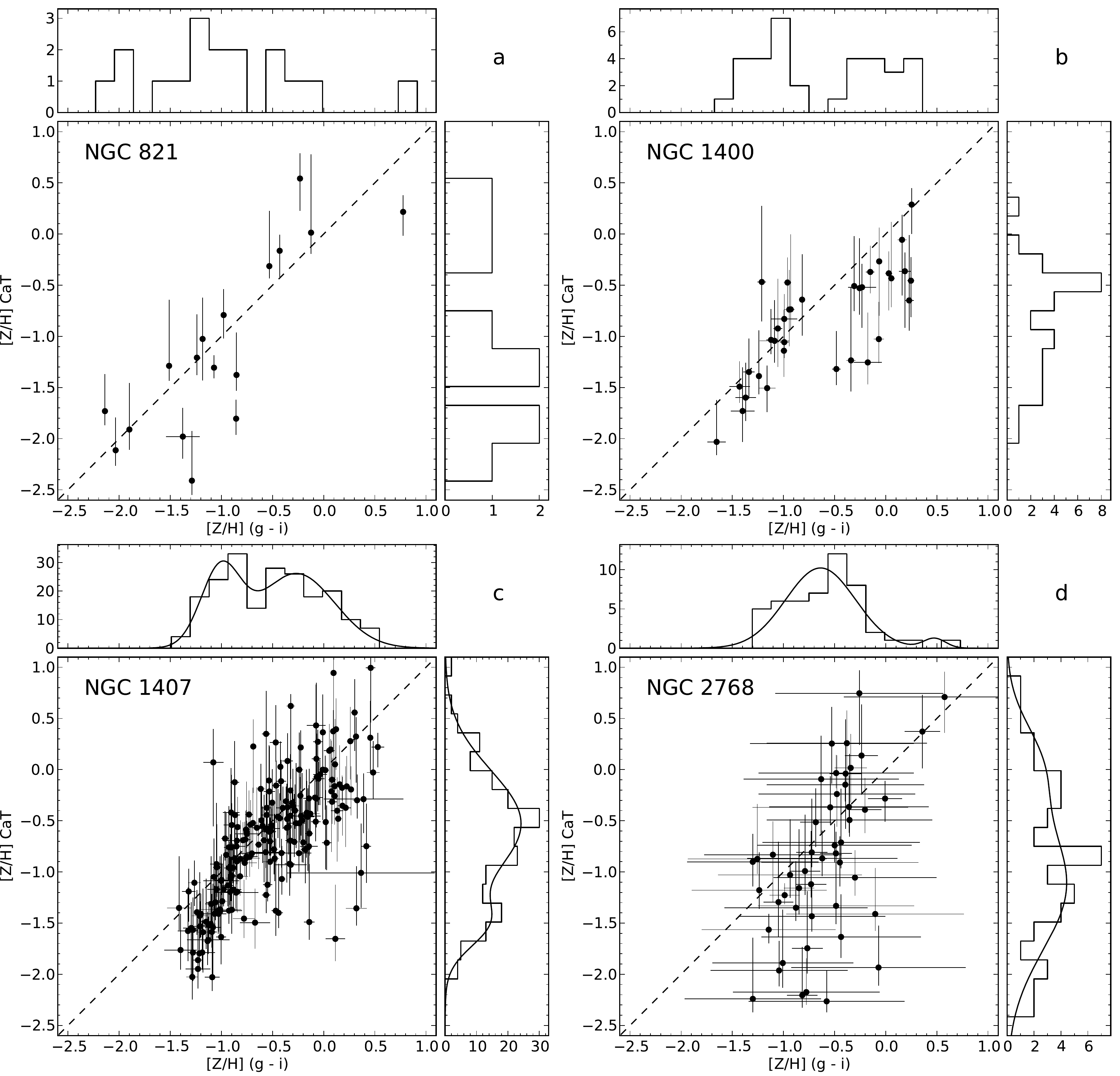}
		\caption{\label{fig:metalmetal}Comparison of colour and CaT-based metallicities for each galaxy.
On the vertical axis is the CaT-based metallicity; the horizontal axis is colour-based metallicity.
The plots on top are the colour-based metallicity histograms; to the right are the CaT-based metallicity histograms.
The bimodal GMM fits, when available, are plotted over the histograms.
The dashed line is a one-to-one line.}

	\end{center}
\end{figure*}

\addtocounter{figure}{-1}
\begin{figure*}
	\begin{center}
		\includegraphics[width=504pt]{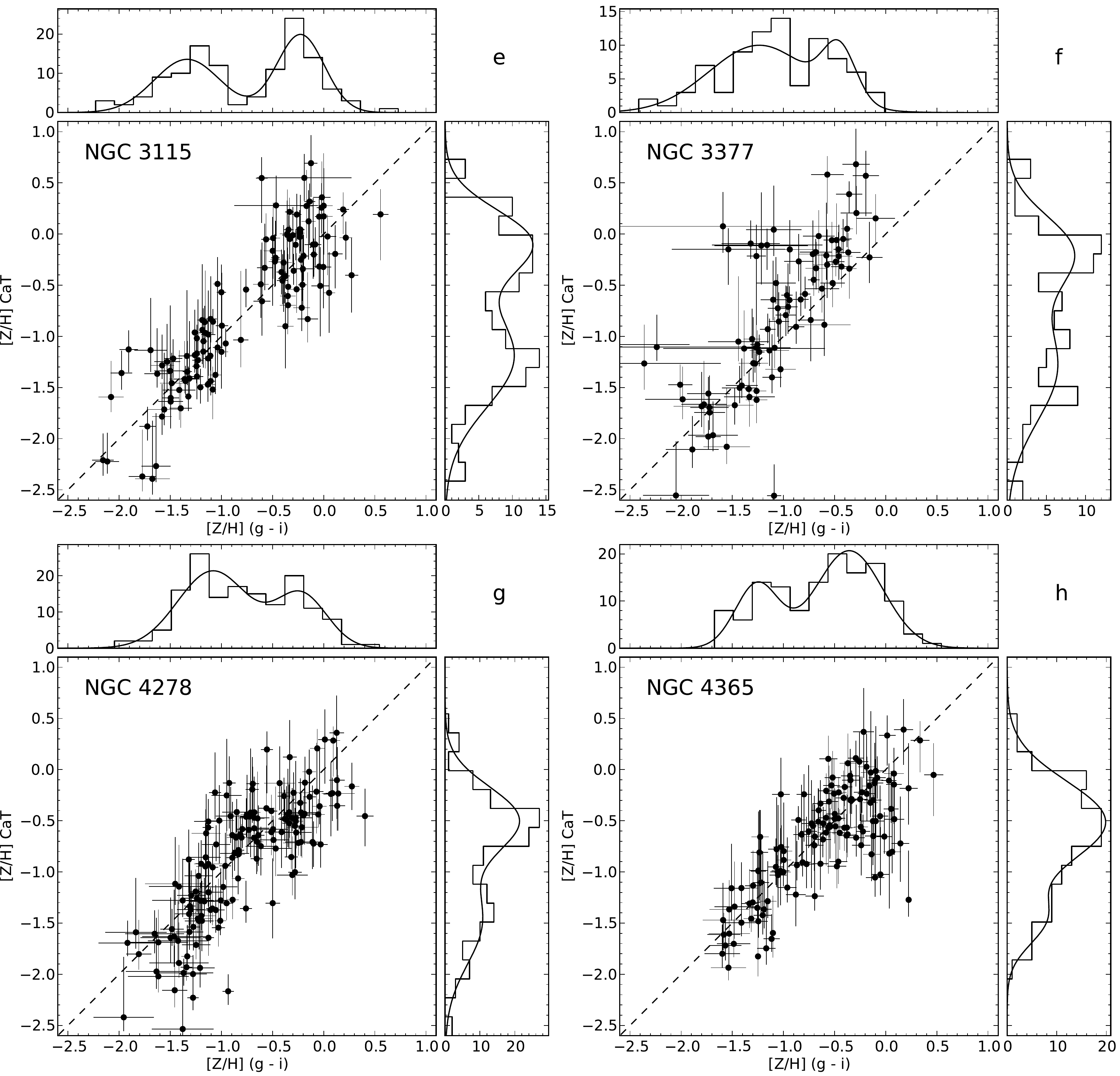}
		\caption{\emph{Continued}}

	\end{center}
\end{figure*}

\addtocounter{figure}{-1}
\begin{figure*}
	\begin{center}
		\includegraphics[width=504pt]{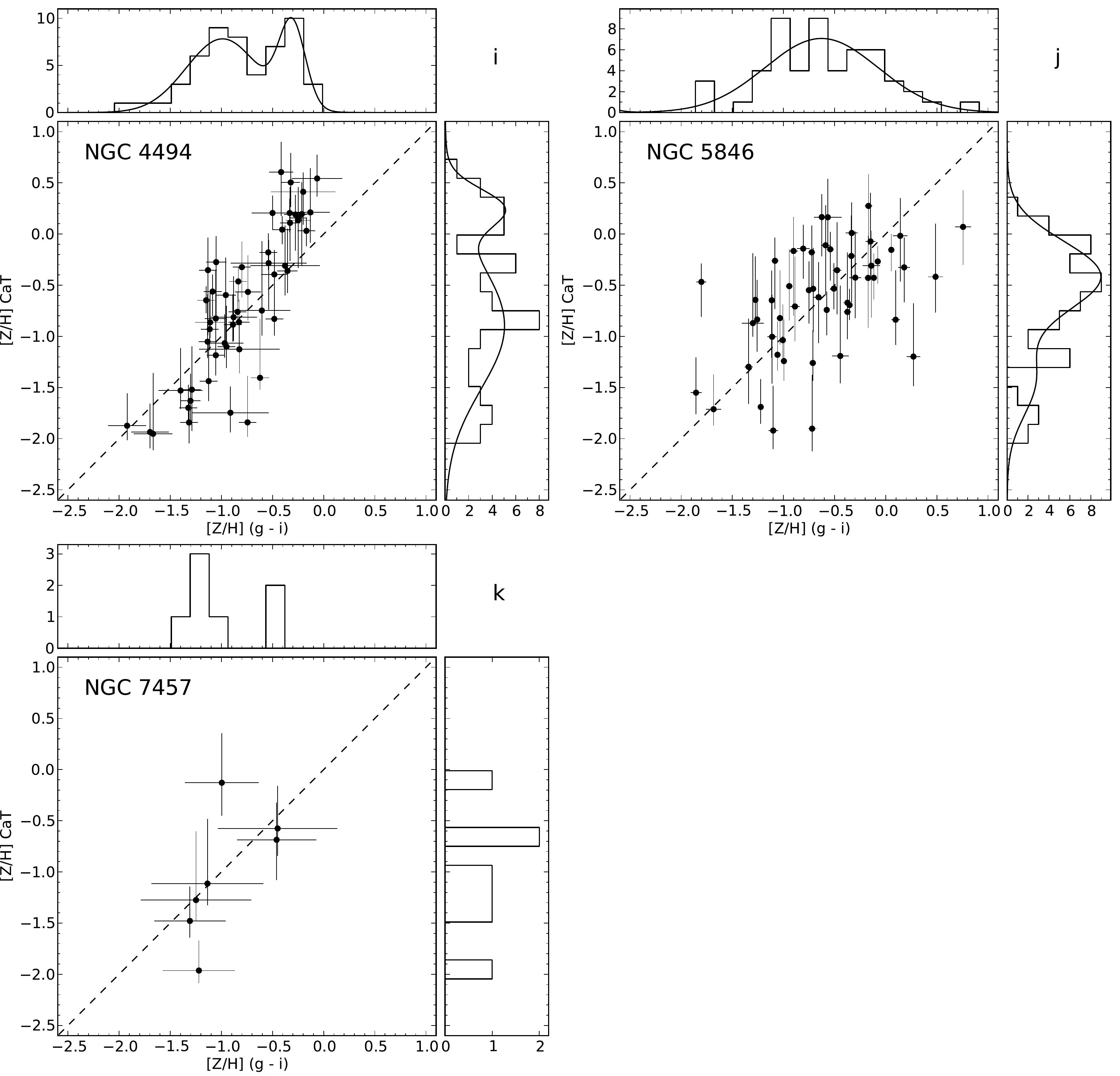}
		\caption{\emph{Continued}}

	\end{center}
\end{figure*}

\section{Discussion}

\subsection{The calcium triplet as a metallicity indicator}
Figure~\ref{fig:compareLickCaT} shows that metallicities measured using the CaT calibrated with SSP models agree with metallicities measured using traditional Lick analysis better than the empirical, \citetalias{1988AJ.....96...92A}-based, calibration.
The agreement is seen over the range of observed metallicities and over a range of both GC magnitudes and galaxy masses.
Since the CaT technique relies on the strength of a single spectral feature, using the CaT to determine metallicities assumes that the age--metallicity, metallicity--calcium abundance and metallicity--horizontal branch morphology relationships are the same for all GCs in all galaxies as well as the same IMF.
The CaT shares this issue with using colours or indices such the [MgFe]$'$ index \citep{2003MNRAS.339..897T} to derive metallicities.

Unlike \citetalias{2003MNRAS.340.1317V} we do not see the CaT saturate with metallicity, at least for metallicities up to solar.
We speculate that this is due to the effect of weak metal lines,
some of which lie in the continuum pass bands of the CaT indices of \citetalias{1988AJ.....96...92A} and \citet{2001MNRAS.326..959C}, acting to lower the pseudo-continuum of the CaT at high metallicities.
Since the strength of the CaT increases and the pseudo-continuum decreases with metallicity, these indices saturate at high metallicity.
As our technique fits the continuum, avoiding regions with weak metal lines, our continuum is less effected and our index does not saturate.

\subsection{The colour--metallicity relation of globular clusters}
\label{colourmetal}
In Figure~\ref{fig:compareLickColour} we compared our empirical colour--metallicity relationship (Equation~\ref{eq:colourmetal}) to those from literature (middle panel) and to a range of different SSP models (lower panel).
Comparing different relations is complicated due to the different metallicity scales used by various authors.
We compared the following empirical relations:
\begin{itemize}
\item \textbf{The \citet{2011MNRAS.416..155F} relation} was based on GMOS photometry and spectroscopy of GCs in NGC 3379, NGC 3923 and NGC 4649.
The metallicities were derived using $\chi^{2}$ minimisation of Lick indices and the models of \citetalias{2003MNRAS.339..897T}. 

\item \textbf{The \citet{2010AJ....140.2101S} relation} used photometry of NGC 5128 obtained using the Y4KCam imager on the Yale/SMARTS 1.0 metre telescope and metallicities derived using the strength of the [MgFe]$'$ index calibrated to Milky Way GCs tied to the Harris metallicity scale.
This relation was converted to the \citetalias{2003MNRAS.339..897T} metallicity scale using Equation~\ref{eq:harrismetal}.

\item \textbf{The \citet{2006ApJ...639...95P} relation} used ACS and SDSS photometry of M49 and M87 GCs together with unpublished photometry of Milky Way GCs.
Metallicities for M49 \citep{2003ApJ...592..866C} and M87 \citep{1998ApJ...496..808C} were derived using the models of \citet{1994ApJS...95..107W} and a limited number of Lick indices while Milky Way metallicities came from the \citet{1996AJ....112.1487H} catalogue.
The Peng et al. relation was converted from $(g-z)$ to $(g-i)$ using Equation~\ref{eq:4365cal} and from the Harris metallicity scale to our metallicity scale using Equation~\ref{eq:harrismetal}.
\end{itemize}

Although they disagree below $(g - i) \sim 0.8$, all three relations agree with our empirical colour--metallicity relations within the uncertainties of each relation.

We also compared our empirical colour--metallicity relation with the SSP models of  \citet{2010MNRAS.404.1639V}, \citet{2005MNRAS.362..799M}, \citet{2011ApJ...743..150Y}, \citet{2009ApJ...699..486C} and \citet{2005AJ....130.2625R}.
We choose to compare the models at the same age ($\sim 13$ Gyr).
The metallicity scale of the models is closer to the metallicity scale of \citet{2009A&A...508..695C} than the \citetalias{2003MNRAS.339..897T} used in this paper so we used Equation~\ref{eq:carrettametal} to convert the model metallicities to the scale of \citetalias{2003MNRAS.339..897T}.
Below we briefly describe each set of models.

\begin{itemize}
\item \textbf{Version 9.1 of the SSP models of \citet{2010MNRAS.404.1639V}.}
We choose models with a \citet{2001MNRAS.322..231K} IMF and with an age of 12.6 Gyr.
Since these models only provide predictions in Johnson--Cousins filters, we use the transformation of \citet{2006A&A...460..339J} to convert $(V - I)$ to $(g - i)$.
For the most metal poor GCs these models predict colours that are too blue while for metal rich GCs they predict colours that are too red.
\\
\item \textbf{The \citet{2005MNRAS.362..799M} SSP models.}
We choose models with a \citet{2001MNRAS.322..231K} IMF and with an age of 13 Gyr.
The Maraston models give a choice of blue or red horizontal branches at low metallicities; we choose the blue models as they provide better agreement with the data.
Except for the most metal poor GCs these models predict redder colours at a given metallicity than most of the observed GCs.
\\
\item \textbf{Version 1.0 of unpublished Yonsei Evolutionary Population Synthesis (YEPS) models.}
These models are described in \citet{2006Sci...311.1129Y} and \citet{2011ApJ...743..150Y}.
Created to study the integrated colours of GCs, these models assume a \citet{1955ApJ...121..161S} IMF and provide both scaled solar and $\alpha$ element enhanced models.
We used models with [$\alpha$/Fe] = 0.3 and an age of 13 Gyr.
Although these models predict realistic colours at intermediate metallicities, they predict colours that are too red at high and low metallicities.
\\
\item \textbf{Version 2.3 of the Flexible Stellar Population Synthesis (FSPS) models of \citet{2009ApJ...699..486C}.}
We used 12.6 Gyr models and a \citet{2001MNRAS.322..231K} IMF.
The FSPS models show good agreement with observations at all metallicities except at the metal rich end, where they are too red.
\\
\item \textbf{The Teramo SPoT models of \citet{2005AJ....130.2625R}.}
These models were used in the \citet{2007ApJ...669..982C} and \citet{2010ApJ...710...51B} theoretical studies of GC bimodality.
These models assume a \citet{1998ASPC..142..201S} IMF and we used their models with an age of 13 Gyr.
We converted the $(V-I)$ colours provided by these models to $(g-i)$ using the transform of \citet{2006A&A...460..339J}.
These models only provide colours similar to observations at intermediate metallicities; at low and high metallicities the predicted colours are too red.
\end{itemize}

Since we compare models with different adopted IMFs, we used the \citet{2010MNRAS.404.1639V} models to investigate the effects of varying the IMF on the colour--metallicity relation.
Moving from a \citet{2001MNRAS.322..231K} IMF to a \citet{1998ASPC..142..201S} IMF caused the $(g-i)$ colour to become bluer but by less than 0.01 mag.
Moving from a \citet{2001MNRAS.322..231K} IMF to a \citet{1955ApJ...121..161S} IMF caused the $(g-i)$ colour to become redder by 0.03 mag with the shift being independent of metallicity.
IMFs more bottom heavy than \citet{1955ApJ...121..161S} have a larger red ward shift.
Similar results were seen with the FSPS models.
We conclude that the colour--metallicity relation is almost identical for bottom light IMFs such as the \citet{2001MNRAS.322..231K} and the \citet{1998ASPC..142..201S} IMFs while bottom heavy IMFs such as the \citet{1955ApJ...121..161S} IMF have redder colours.
Therefore the SPoT models which use a \citet{1998ASPC..142..201S} IMF is directly comparable to the other models which use a \citet{2001MNRAS.322..231K} IMF while the YEPS models are likely a few hundredths of a magnitude too red due to their choice of a \citet{1955ApJ...121..161S} IMF.

All the models predict redder colours at the metal rich end than are observed. 
It is unknown whether this is due to a failing of the colour--metallicity predictions of SSP models, due to an issue with the \citetalias{2003MNRAS.339..897T} models used to derive the metallicities, or a mixture of the two.
Out of the five SSP models we compare, the FSPS models of \citet{2009ApJ...699..486C} appear to provide the closest match to observations.
However it is difficult to quantitatively show that one model agrees with observations better than the others due to similarities of the models and the limitations of our data. 

In five of our eleven galaxies (NGC 1400, NGC 1407, NGC 2768, NGC 3377 and NGC 4494) the KS test (Figure~\ref{fig:zhist}) provides evidence that the CaT-based metallicities are not drawn from the same distribution as the colour-based metallicities while in a sixth (NGC 4278) the KS test says there is a low but not significant probability that the two metallicities are not drawn from the same distribution.
In the case of NGC 2768 the poor agreement between colour-based and CaT-based metallicity is likely due to the poor photometry available for this galaxy.
The inner regions of NGC 3377 also suffer from poor photometry; however when GCs with poor photometry are removed, the disagreement between colour and CaT-based metallicity remains.

Although NGC 1407 shows good agreement at high metallicities it displays poor agreement at low metallicities with the colour-based metallicities being more metal rich than the CaT-based metallicities (Figures \ref{fig:zhist}$c$, \ref{fig:metalmetal}$c$).
NGC 1407's GCs possess a strong blue tilt \citep{2006ApJ...636...90H}, a colour--magnitude (mass--metallicity) relation where the colour (metallicity) peak of the blue (metal poor) subpopulation becomes redder (more metal rich) with brighter magnitudes (higher mass) \citep{2006AJ....132.2333S, 2006ApJ...653..193M}.
The blue tilt is thought to have caused by GCs self enrichment as more massive GCs are able to hold on to more of the metals formed by the first stars to form in the GCs \citep[e.g.][]{2009ApJ...695.1082B}.
It is plausible that this self enrichment changes the abundance pattern (for example the helium abundance) in such a way that the colour--metallicity relation (or at least the colour--CaT relation) is changed.
The GCs we have CaT measurements for are effected by NGC 1407's blue tilt as can be seen in Figure~\ref{fig:mag}$c$.
NGC 4278 is possibly another example of this effect as it also possesses a blue tilt (Usher et al. in prep.) and, like NGC 1407, shows weak evidence that the metal poor GCs are redder than in other galaxies (Figures \ref{fig:zhist}$g$, \ref{fig:metalmetal}$g$).

Another possible explanation is that the metal poor GCs in these galaxies have a different IMF than metal rich GCs in these galaxies or all GCs in other galaxies.
Using CaT SSP models of \citetalias{2003MNRAS.340.1317V} and the colours from the SSP models of \cite{2010MNRAS.404.1639V} we found that going from a \citet{2001MNRAS.322..231K} IMF to a \citet{1955ApJ...121..161S} IMF increased the measured CaT metallicity by 0.02 dex but make the $(g-i)$ colour 0.03 mag redder which due to the steeper colour--metallicity relation would increase the metallicity derived from metallicity by 0.25 dex while a bottom heavy IMF similar that found by \citet{2010Natur.468..940V} in the most massive ellipticals (slope $x = -3$) increased the measured CaT metallicity by 0.07 dex but would make the colour 0.12 mag redder which would increase the colour based metallicity by 0.9 dex.
Thus the redder colours of metal poor GCs in NGC 1407 and NGC 4278 could be explained by a more bottom heavy IMF for these GCs than that of other GCs.

Both NGC 3377 and NGC 4494 show the opposite trend, with agreement at low metallicities and the colour-based metallicities giving lower values than the CaT-based ones at high metallicities (Figures \ref{fig:zhist}$f, i$, \ref{fig:metalmetal}$f, i$).
In both NGC 3377 and NGC 4494 the relationship between colour and metallicity appears to be linear and to be steeper than in other galaxies.
The bluer colours of metal rich GCs in these galaxies could be explained by some of the metal rich GCs being younger than in other galaxies.
In FSPS models of \citet{2009ApJ...699..486C}, at solar metallicity the $(g - i)$ colour difference between a 7.9 Gyr model and a 13.3 Gyr model is 0.10.
Since the star formation history of early-type galaxies is dependent on mass and environment \citep[e.g.][]{2000AJ....120..165T, 2002MNRAS.330..547T,2003AJ....125.2891C, 2005ApJ...632..137N, 2005ApJ...621..673T, 2012MNRAS.419.3167S} it is natural to expect GCs in galaxies of different masses and environments to have different age--metallicity relations.
Some evidence of this is seen in the Milky Way where GCs associated with dwarf galaxies that have been accreted on to the Milky Way show a different, steeper age--metallicity relation than those thought to have formed within the Galaxy itself \citep{2010MNRAS.404.1203F, 2011ApJ...738...74D}.
NGC 1400 is different again with agreement at low metallicities while having more metal rich colour-based values than CaT values at high metallicities (Figures \ref{fig:zhist}$b$, \ref{fig:metalmetal}$b$).

As noted in Section~\ref{photometry}, our photometry is heterogeneous. 
Different photometric zero points could shift the colour--metallicity relation along the colour axis or change its slope.
We assumed one extinction value for all GCs in each galaxy so any differential reddening or internal extinction could affect the colours of different GCs with in a galaxy to differing degrees.
It is unclear whether our heterogeneous photometry is responsible for the different colour--metallicity relations seen in different galaxies in our sample.
If the colour--metallicity relation does vary from galaxy to galaxy it could be caused by variations in the ages or other stellar population parameters such as the IMF of the GCs between galaxies.

\subsection{Globular cluster metallicity distributions}
\label{metaldistro}
Although spectroscopic studies tend to sample the colour distribution well they are inherently biased to brighter magnitudes.
If the stellar populations of the brightest GCs are different from the majority, then using only the brightest GCs will give a biased result.
The blue tilt causes the metallicity distribution of the brightest GCs to be more metal rich and less bimodal than the metallicity distribution of the fainter GCs.
We also note that the larger relative errors and smaller numbers in spectroscopic studies can make identification of metallicity bimodality more difficult than finding colour bimodality in photometric studies. 

Using GMM (Section~\ref{GMM}, Table~\ref{tab:gmm-metal}, Figure~\ref{fig:metalmetal}), we detected CaT metallicity bimodality with a probability of greater than 0.95 in NGC 3115, NGC 4278 and NGC 4365 and greater than 0.90 in NGC 1407, NGC 5846 and NGC 3377.
NGC 821, NGC 1400 and NGC 7457 do not have enough measured GCs to determine whether their distributions are bimodal or not.
In NGC 2768 we do not see metallicity bimodality nor is the colour distribution of GCs with CaT measurements bimodal.
Due to the poor quality of the photometry for this galaxy it is unclear whether we should expect metallicity bimodality.

\begin{figure}
	\begin{center}
		\includegraphics[width=240pt]{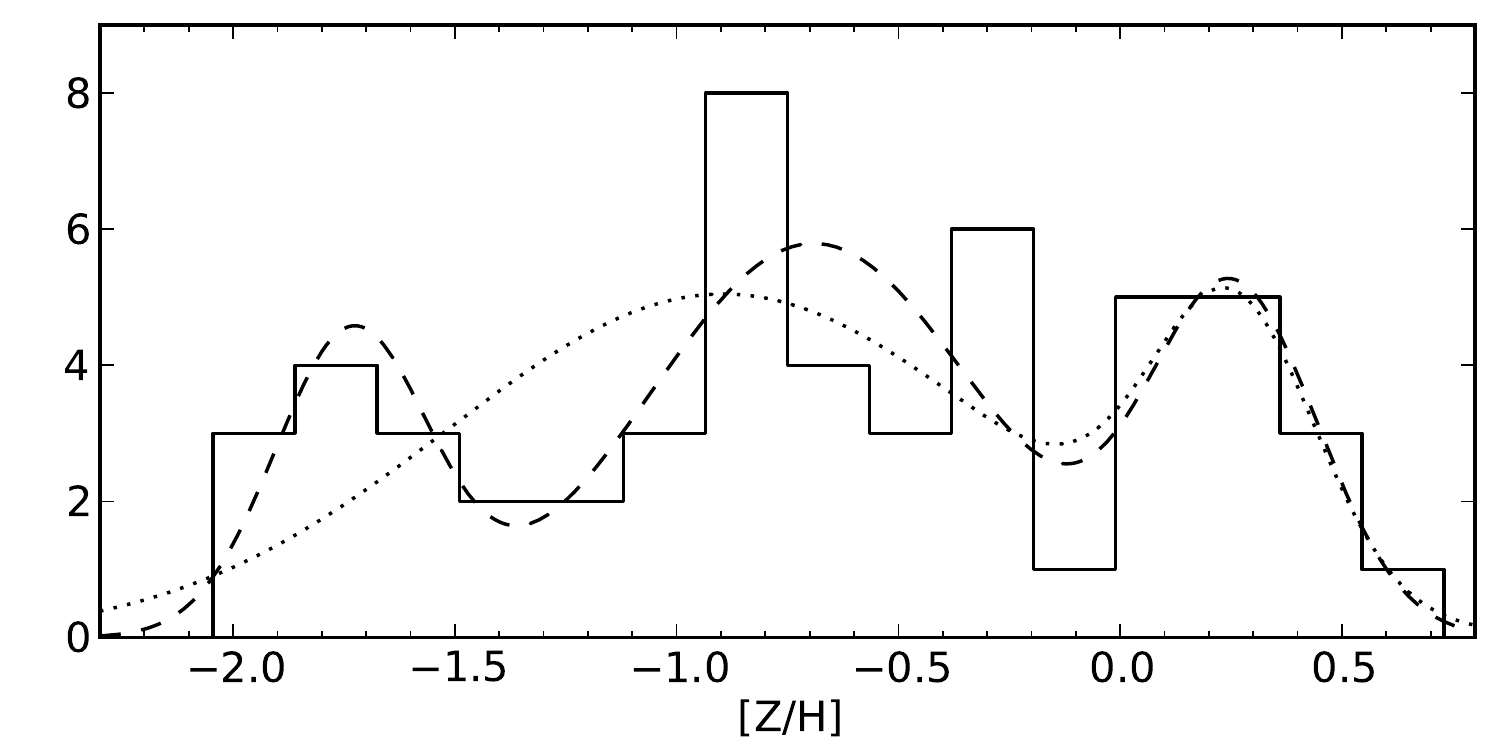}
		\caption{\label{fig:ngc4494tri} CaT metallicity distribution for NGC 4494.
The solid line is a histogram of the measured metallicities.
The dotted line is the two peak GMM fit while the dashed line is the three peak GMM fit.
GMM returns a higher probability that three Gaussians are preferred over one ($p = 0.90$) than two Gaussians ($p = 0.82$) are preferred over one.
This hints that NGC 4494's metallicity distribution is trimodal rather than unimodal or bimodal.}

	\end{center}
\end{figure}

NGC 4494 presents an interesting case.
Although it does not meet one of our criteria for bimodality (a probability that two Gaussians are preferred over one of at least 0.9), GMM still gives a high probability that a two Gaussians are preferred over one of 0.819.
The colour distribution of GCs with CaT measurements is bimodal (Table~\ref{tab:gmm-colour}) so we would expect metallicity bimodality.
Since the metallicity distribution of NGC 4494 appears trimodal (Figure~\ref{fig:ngc4494tri}) we used GMM to test where the metallicity distribution is trimodal.
We found peaks at [Z/H] $= -1.73 \pm 0.19$, $-0.69 \pm 0.14$ and $0.26 \pm 0.11$, widths of $0.17 \pm 0.12$, $0.38 \pm 0.12$ and $0.19 \pm 0.07$ and $10.5 \pm 5.8$, $29.5 \pm 7.7$ and $13.0 \pm 5.0$ GCs in each peak.
GMM gives a probability of 0.895 that a trimodal distribution is preferred over a unimodal distribution, a higher probability than for bimodal over unimodal.
The peaks are well separated ($D = 3.55$ and $D = 2.32$) and the kurtosis is negative $k = -1.08$ supporting the existence of trimodality.
The bimodal and trimodal GMM fits are plotted with the NGC 4494 metallicity histogram in Figure~\ref{fig:ngc4494tri}.
Due to the low number (53) of GCs these results should be treated with caution. 
\citetalias{2011MNRAS.415.3393F} found evidence for three kinematic subpopulations in that although the bluest GCs and the reddest GCs in NGC 4494 show evidence for rotation, the intermediate colour GCs do not rotate.
Taken together these lines of evidence suggest, but do not prove, that NGC 4494 has three GC subpopulations.

NGC 4365 shows evidence for colour trimodality \citep{2012MNRAS.420...37B} and shows different kinematics for the different GC colour subpopulations \citep{Christina2012} so we also ran GMM with three peaks on this galaxy.
Although GMM finds peaks at [Z/H] $= -1.53$, $-0.70$ and $-0.20$ which are located at the same places as the colour peaks found by \citet{2012MNRAS.420...37B} transformed into metallicity, GMM does not find three Gaussians to be a significant improvement over one ($p = 0.671$).  
Due to the smaller numbers and larger errors of the CaT metallicities compared to the colours, it is not surprising that trimodality was not found to be an improvement.

In general the colour distributions of the GCs with CaT measurements are similar to the photometric GC candidate colour distribution in eight galaxies with sufficient numbers for GMM to be run.
However the blue peak of the GCs with CaT measurements in NGC 1407 is redder than the candidate blue peak in this galaxy due to the blue tilt while the red peak of the GCs in NGC 3377 is redder than that of the photometric GC candidates.
The absolute magnitude limit of the GCs with CaT measurements varies dramatically as can be seen in the colour magnitude diagrams in Figure~\ref{fig:mag}.
In NGC 821, NGC 1400, NGC 1407 and NGC 5846, among the most distant galaxies in this study, the majority of GCs with CaT measurements are brighter than $M_i = -10$ while in NGC 3115 and NGC 3377, the closest galaxies, GCs as faint as $M_i = -8$ have CaT metallicities.

\begin{figure}
	\begin{center}
		\includegraphics[width=240pt]{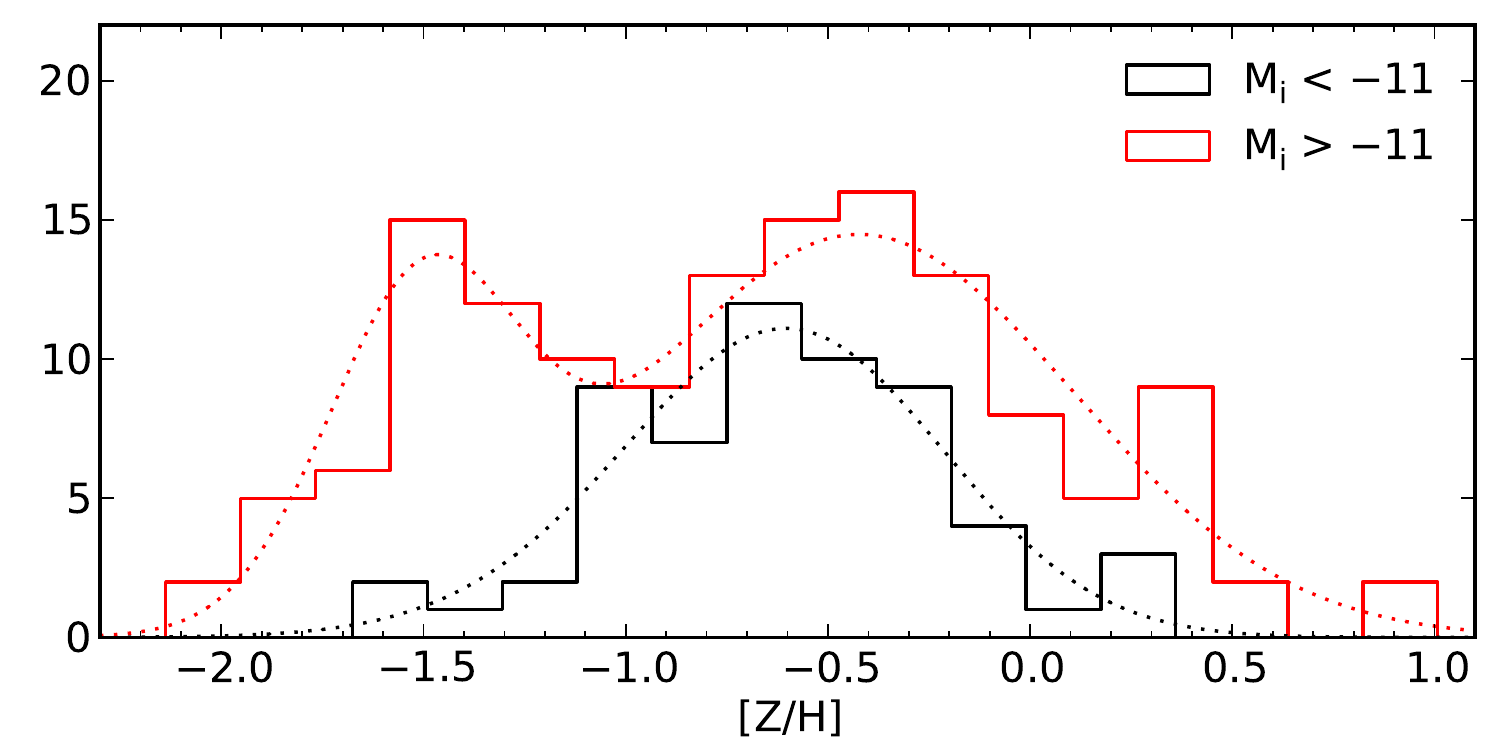}
		\caption{\label{fig:ngc1407bi} The CaT metallicity distribution of NGC 1407 split by magnitude.
The black (red) histogram shows the distribution of GCs brighter (fainter) than $M_{i} = -11$ while the dotted lines give the GMM fits.
For the bright GCs GMM gives a low probability that two Gaussians are preferred over one ($p = 0.014$) while for the bright GCs GMM gives a high probability ($p = 0.996$).}

	\end{center}
\end{figure}

In Figure~\ref{fig:mag}$c$ where the colour magnitude and metallicity magnitude diagrams of NGC 1407 are shown, the colour and metallicity distributions of GCs brighter than $M_{i} = -11$ appear different from those of fainter GCs.
We split the GCs with CaT measurements in NGC 1407 into two magnitude bins and ran GMM on each.
The results are plotted in Figure~\ref{fig:ngc1407bi}.
For GCs brighter than $M_{i} = -11$ the distribution looks quite unimodal.
Fainter than $M_{i} = -11$ the distribution looks bimodal.
GMM confirmed this with the bright population showed a low probability that 2 Gaussians are preferred to 1 (0.137) while the faint population showed high probability that 2 Gaussians are preferred (0.996).
If only the 100 brightest GCs in NGC 1407 had been studied spectroscopically bimodality would not have likely been found.
This may explain why the \citet{1998ApJ...496..808C} study of M87's GCs did not produce convincing proof of metallicity bimodality.
It is known that in several galaxies including NGC 1407 \citep{2009AJ....137.4956R}, M87 \citep{2011ApJS..197...33S} and NGC 5128 \citep{2010AJ....139.1871W} GCs brighter than $M_{i} \sim -11$ have different kinematics than fainter GCs further supporting the idea that the brightest GCs form a different population.

Out of the eight galaxies with significant numbers of GCs with CaT metallicity measurements, we see bimodality in six and evidence for trimodality in another.
In the eighth galaxy we do not see multimodality. 
However due to the poor quality of the photometry in this galaxy it is unclear whether we would expect bimodality based on the colour distribution of the observed GCs.
Metallicity bimodality (or multi-modality) appears to be a common feature in massive early-type galaxies. 

\subsection{Comparison with previous calcium triplet measurements}
\label{previous}
\citetalias{2010AJ....139.1566F} used the CaT to measure the metallicity of 144 GCs in NGC 1407 while \citetalias{2011MNRAS.415.3393F} used the same technique to measure the metallicity of 54 GCs in NGC 4494. 
In this study we have added additional GCs to the NGC 1407 data set and remeasured the CaT for all GCs in NGC 1407 and NGC 4494.
Unlike the previous studies we normalised the continuum with out human intervention (Section~\ref{caterrors}).
Rather than calculate the uncertainty on the CaT measurements using the S/N as was done in \citetalias{2010AJ....139.1566F} and \citetalias{2011MNRAS.415.3393F} we used Monte Carlo resampling to calculate the errors which results in larger uncertainties.
Unlike the earlier studies we corrected for an apparent S/N bias in the CaT index measurement (Section~\ref{bias}).
We used a CaT metallicity calibration based on the SSP models of \citetalias{2003MNRAS.340.1317V} rather than an empirical calibration based on the strength of the CaT in Milky Way GCs as measured by \citetalias{2003MNRAS.340.1317V} (Section~\ref{calib}).

Despite the somewhat different analysis we get qualitatively similar results.
The qualitative distributions of CaT indices in \citetalias{2010AJ....139.1566F} and \citetalias{2011MNRAS.415.3393F} both agree with this work, with NGC 1407 showing a bimodal distribution and NGC 4494 not showing clear bimodality.
The shapes of the colour--CaT index relations we find are similar to those of the previous studies with NGC 1407 showing the relation steepening at low metallicities and NGC 4494 showing a linear relation.

\begin{figure}
	\begin{center}
		\includegraphics[width=240pt]{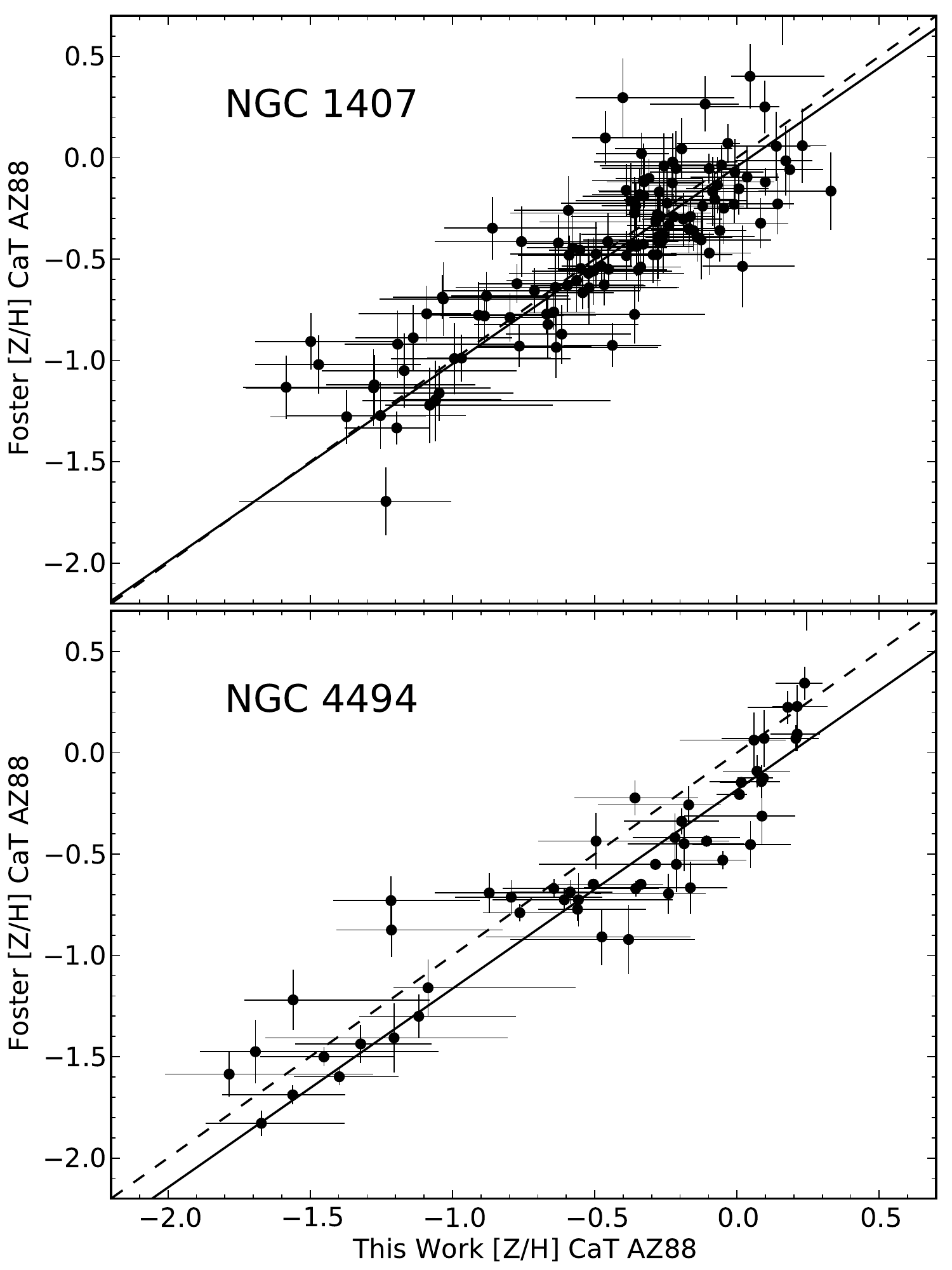}
		\caption{\emph{Top} Comparison of CaT metallicities in NGC 1407 from \citet{2010AJ....139.1566F} with this work.
\emph{Bottom} Comparison of CaT metallicities in NGC 4494 from \citet{2011MNRAS.415.3393F}  with this work.
Over plotted with solid lines are linear fits to the metallicities while the dashed line is one-to-one.
Both studies give consistent metallicities for NGC 1407.
While the slope is consistent with unity, the NGC 4494 metallicities are 0.2 dex lower than those of this work.
For consistency we use the \citet{1988AJ.....96...92A}-based calibration (Equation~\ref{eq:cal}) rather than single stellar population model metallicity calibration (Equation~\ref{eq:catfecal}) used elsewhere in this work and convert the metallicities of \citet{2010AJ....139.1566F} and \citet{2011MNRAS.415.3393F} to our metallicity scale. \label{fig:compareFoster}}

	\end{center}
\end{figure}

In Figure~\ref{fig:compareFoster} we compare our measurements to those of \citetalias{2010AJ....139.1566F} and \citetalias{2011MNRAS.415.3393F}. 
To be consistent with the calibration technique used in these papers we use the \citetalias{1988AJ.....96...92A} metallicity calibration (Equation~\ref{eq:cal}), rather than the SSP model based calibration (Equation~\ref{eq:catfecal}), for this comparison only.
Additionally we use Equation~\ref{eq:catanzfecal} in place of equation 2 of \citetalias{2010AJ....139.1566F} to convert the CaT values of \citetalias{2010AJ....139.1566F} and \citetalias{2011MNRAS.415.3393F} into metallicities so any comparison is not biased by choice of metallicity scale.
Although both measurements for NGC 1407 are consistent, for NGC 4494 we find there is a significant offset with our measurements having metallicities 0.2 dex lower on average than those of \citetalias{2011MNRAS.415.3393F}.

Unlike \citetalias{2010AJ....139.1566F} we do not see significant Paschen lines in any of our fitted NGC 1407 spectra.
We note that the strength of the Paschen line index used increases with metallicity due to the presence of weak metal lines in the index passband so most of the higher Paschen index GCs tend to be high metallicity GCs.
\citet{Jean2012} used SSP models to show that horizontal branch morphology has little effect on the strength of the CaT.
Like \citetalias{2010AJ....139.1566F} we find that the brightest GCs in NGC 1407 have similar CaT index values.
However we find that the colours of the brightest GCs when transformed into metallicity are also similar.

\section{Conclusions}
We have used the strength of the calcium triplet (CaT) to measure the metallicities of 903 globular clusters (GCs) in 11 galaxies ranging from brightest group ellipticals to isolated lenticulars.
This is the largest sample of spectroscopic GC metallicities ever assembled and the first to provide large numbers ($> 50$) of metallicities for multiple galaxies.
We showed that the CaT is a viable method of measuring the metallicity of extragalactic GCs with our CaT derived metallicities agreeing with literature Lick index-based metallicities.

Using literature metallicities we defined a new colour--metallicity relation (Equation~\ref{eq:colourmetal}).
Although agreement is seen between the metallicities predicted by this relation and CaT-based metallicities in half of our galaxies, the remaining galaxies show some disagreement either at low metallicity or at high metallicity.
It is unclear whether these differences are caused by our heterogeneous photometry or real differences in the colour--metallicity relation between galaxies.
Differences in the colour--metallicity relation could be driven by differences in the age or the IMF of GCs between galaxies.

We found spectroscopic metallicity bimodality in six of the eight galaxies with more than forty GC measurements, while evidence of trimodality was seen in one of the remaining galaxies (NGC 4494) and we can not rule out bimodality in the other (NGC 2768).
We also found that the brightest GCs in NGC 1407 posses a different, unimodal metallicity distribution than the fainter, bimodal GCs.
We caution that studies that rely on the properties of the brightest GCs to infer the properties of the GC system may give misleading results.
These bimodality results are similar to what has been seen in previous spectroscopic studies of the GC metallicity distributions of early-type galaxies where three galaxies (NGC 5128, \citealt{2008MNRAS.386.1443B}, NGC 4594, \citealt{2011MNRAS.417.1823A} and M49, \citealt{2007AJ....133.2015S}) show evidence of bimodality while the situation in a fourth galaxy \citep[M87][]{1998ApJ...496..808C} is inconclusive.
When the previous studies are combined with our own results, nine of the twelve galaxies show evidence for bimodality while in the remaining three galaxies bimodality can not be ruled out.

The spatial distributions and kinematics of GCs system also favour multiple, distinct subpopulations.
Red GCs are more centrally concentrated than blue GCs \citep[e.g.][]{1996AJ....111.1529G, 2006MNRAS.367..156B, 2009ApJ...703..939H, 2011MNRAS.416..155F, Duncan2012}.
In addition the colour subpopulations often show different kinematics \citep{2003ApJ...591..850C, 2010A&A...513A..52S, 2010ApJ...709.1083L, 2011ApJS..197...33S, Vince2012} with sharp changes in rotation occurring at the blue/red dividing line. 
NIR photometric studies \citep[e.g.][]{2007ApJ...660L.109K, 2008MNRAS.389.1150S, 2012A&A...539A..54C} also see evidence for bimodality in several galaxies although there are unresolved issues with the connection between metallicity and NIR photometry \citep{2012ApJ...746...88B}.

Our observed GC metallicity distributions, together with other observations of GC systems, favour a picture where most massive galaxies have bimodal GC systems.
Since GC metallicity bimodality appears to be common most early-type galaxies should have experienced two periods of intense star formation.

\begin{table*}
\caption{Calcium Triplet Measurements and Metallicities}
\label{tab:data}
\begin{tabular}{l c c c c c c}
Name & RA. & Dec. & $(g - i)$ & $i$ & CaT & [Z/H] \\
     & [deg] & [deg] & [mag] & [mag] & [\AA] & [dex] \\ 
(1) & (2) & (3) & (4) & (5) & (6) & (7) \\ \hline
NGC4494\_GC100 & 187.811375 & 25.779614 & $0.93 \pm 0.03$ & $21.04 \pm 0.02$ & $7.44_{-0.64}^{+0.54}$ & $-0.32_{-0.29}^{+0.25}$ \\ 
NGC4494\_GC101 & 187.855092 & 25.783281 & $0.98 \pm 0.08$ & $20.81 \pm 0.06$ & $6.52_{-0.54}^{+1.47}$ & $-0.75_{-0.25}^{+0.68}$ \\ 
NGC4494\_GC102 & 187.873400 & 25.783786 & $1.04 \pm 0.03$ & $21.84 \pm 0.02$ & $9.46_{-0.94}^{+0.64}$ & $0.60_{-0.43}^{+0.30}$ \\ 
\ldots & \ldots & \ldots & \ldots & \ldots & \ldots & \ldots \\ \hline
\end{tabular}

\medskip
The full version of this table is provided in a machine readable form in the online Supporting Infromation.
\emph{Notes} Column (1): Globular cluster (GC) IDs.
The IDs are the same as in \citet{Vince2012} except for the GCs in NGC 4494 where 'NGC4494\_' has been appended to the IDs from \citetalias{2011MNRAS.415.3393F}.
Column (2) and (3): Right ascension and declination in the J2000.0 epoch, respectively.
Column (4): Adopted $(g - i)$ colour.
Column (5): Adopted $i$ magnitude.
Column (6): CaT index measurement corrected for S/N bias.
Column (7): Total metallicity.
\end{table*}

\begin{table*}\caption{Bimodal Metallicity Distribution Results using GMM}\label{tab:gmm-metal}
\begin{center}
\begin{tabular}{l c c c c c c c c c c c}
Galaxy & Sample & $N$ & $\mu_{poor}$ & $\sigma_{poor}$ & $\mu_{rich}$ & $\sigma_{rich}$ & $f_{poor}$ & $p$ & $D$ & $k$ & Bi \\
(1) & (2) & (3) & (4) & (5) & (6) & (7) & (8) & (9) & (10) & (11) & (12) \\ \hline
NGC 1407 & $(g-i)$ & 202 & $-1.01 \pm 0.08$ &$0.19 \pm 0.04$ & $-0.27 \pm 0.10$ &$0.38 \pm 0.05$ & $0.34 \pm 0.11$ & 0.999 & $2.50 \pm 0.28$ & -0.96 & Y \\
 & CaT & 202 & $-1.52 \pm 0.26$ &$0.23 \pm 0.12$ & $-0.53 \pm 0.36$ &$0.52 \pm 0.11$ & $0.16 \pm 0.22$ & 0.946 & $2.46 \pm 0.66$ & -0.40 & Y \\
\hline
NGC 2768 & $(g-i)$ & 49 & $-0.64 \pm 0.28$ &$0.34 \pm 0.10$ & $0.48 \pm 0.48$ &$0.11 \pm 0.14$ & $0.96 \pm 0.36$ & 0.645 & $4.40 \pm 1.51$ & 0.37 & N \\
 & CaT & 49 & $-1.12 \pm 0.44$ &$0.64 \pm 0.20$ & $0.07 \pm 0.43$ &$0.40 \pm 0.17$ & $0.78 \pm 0.29$ & 0.141 & $2.24 \pm 0.68$ & -0.78 & N \\
\hline
NGC 3115 & $(g-i)$ & 122 & $-1.33 \pm 0.04$ &$0.32 \pm 0.04$ & $-0.23 \pm 0.03$ &$0.23 \pm 0.03$ & $0.49 \pm 0.05$ & 0.999 & $3.94 \pm 0.37$ & -1.06 & Y \\
 & CaT & 122 & $-1.19 \pm 0.10$ &$0.50 \pm 0.09$ & $-0.07 \pm 0.06$ &$0.32 \pm 0.05$ & $0.57 \pm 0.08$ & 0.998 & $2.67 \pm 0.50$ & -0.75 & Y \\
\hline
NGC 3377 & $(g-i)$ & 84 & $-1.24 \pm 0.41$ &$0.49 \pm 0.12$ & $-0.45 \pm 0.24$ &$0.16 \pm 0.12$ & $0.80 \pm 0.27$ & 0.903 & $2.13 \pm 0.72$ & -0.22 & Y \\
 & CaT & 84 & $-1.29 \pm 0.29$ &$0.54 \pm 0.15$ & $-0.16 \pm 0.15$ &$0.35 \pm 0.12$ & $0.56 \pm 0.17$ & 0.910 & $2.49 \pm 0.62$ & -0.66 & Y \\
\hline
NGC 4278 & $(g-i)$ & 150 & $-1.08 \pm 0.07$ &$0.34 \pm 0.05$ & $-0.22 \pm 0.08$ &$0.25 \pm 0.05$ & $0.66 \pm 0.09$ & 0.995 & $2.86 \pm 0.37$ & -0.82 & Y \\
 & CaT & 150 & $-1.48 \pm 0.15$ &$0.42 \pm 0.09$ & $-0.48 \pm 0.05$ &$0.32 \pm 0.06$ & $0.41 \pm 0.11$ & 0.999 & $2.65 \pm 0.53$ & -0.47 & Y \\
\hline
NGC 4365 & $(g-i)$ & 131 & $-1.26 \pm 0.07$ &$0.21 \pm 0.04$ & $-0.36 \pm 0.06$ &$0.33 \pm 0.04$ & $0.30 \pm 0.06$ & 0.999 & $3.26 \pm 0.31$ & -0.97 & Y \\
 & CaT & 131 & $-1.47 \pm 0.23$ &$0.25 \pm 0.10$ & $-0.51 \pm 0.12$ &$0.40 \pm 0.07$ & $0.18 \pm 0.16$ & 0.951 & $2.88 \pm 0.53$ & -0.58 & Y \\
\hline
NGC 4494 & $(g-i)$ & 53 & $-0.99 \pm 0.18$ &$0.35 \pm 0.08$ & $-0.31 \pm 0.12$ &$0.13 \pm 0.05$ & $0.71 \pm 0.16$ & 0.954 & $2.55 \pm 0.67$ & -0.58 & Y \\
 & CaT & 53 & $-0.90 \pm 0.38$ &$0.62 \pm 0.19$ & $0.26 \pm 0.32$ &$0.19 \pm 0.17$ & $0.80 \pm 0.27$ & 0.819 & $2.51 \pm 0.53$ & -1.08 & N \\
\hline
NGC 5846 & $(g-i)$ & 54 & $-2.79 \pm 0.91$ &$0.14 \pm 0.32$ & $-0.63 \pm 0.33$ &$0.55 \pm 0.16$ & $0.02 \pm 0.30$ & 0.841 & $5.37 \pm 2.20$ & 1.19 & N \\
 & CaT & 54 & $-1.42 \pm 0.31$ &$0.35 \pm 0.14$ & $-0.41 \pm 0.13$ &$0.34 \pm 0.08$ & $0.24 \pm 0.18$ & 0.925 & $2.93 \pm 0.99$ & -0.43 & Y \\
\hline
\end{tabular}

\medskip
\emph{Notes} Column (1): Galaxy name.
Column (2): Colour or CaT-based metallicity.
Column (3): Number of GCs with CaT measured.
Column (4): Mean metallicity of the metal poor subpopulation.
Column (5): Dispersion of the metal poor subpopulation.
Column (6): Mean metallicity of the metal rich subpopulation.
Column (7): Dispersion of the metal rich subpopulation.
Column (8): Fraction of the clusters in the metal poor population.
Column (9): p-value that a bimodal fit is preferred over a unimodal fit.
Column (10): Separation of the GMM peaks normalised by their width.
Column (11): The kurtosis of the sample.
Column (12): Bimodality of the sample. A p-value greater than 0.9, a separation of $D > 2$ and a negative kurtosis are all required for a sample to be considered bimodal. 
We did not run GMM on NGC 821 (17 GCs), NGC 1400 (34 GCs) and NGC 7457 (7 GCs) due to the low number of GCs with CaT measurements in these galaxies. 
\end{center}
\end{table*}

\begin{table*}\caption{Bimodal Colour Distribution Results using GMM}\label{tab:gmm-colour}
\begin{center}
\begin{tabular}{l c c c c c c c c c c c}
Galaxy & Sample & $N$ & $\mu_{blue}$ & $\sigma_{blue}$ & $\mu_{red}$ & $\sigma_{red}$ & $f_{blue}$ & $p$ & $D$ & $k$ & Bi \\
(1) & (2) & (3) & (4) & (5) & (6) & (7) & (8) & (9) & (10) & (11) & (12) \\ \hline
NGC 821 & All & 110 & $0.81 \pm 0.05$ &$0.04 \pm 0.03$ & $1.03 \pm 0.10$ &$0.16 \pm 0.04$ & $0.18 \pm 0.26$ & 0.969 & $1.87 \pm 0.70$ & -0.48 & N \\
\hline
NGC 1400 & All & 1010 & $0.80 \pm 0.01$ &$0.06 \pm 0.01$ & $1.09 \pm 0.02$ &$0.10 \pm 0.01$ & $0.53 \pm 0.05$ & 0.990 & $3.53 \pm 0.37$ & -1.16 & Y \\
\hline
NGC 1407 & All & 3312 & $0.82 \pm 0.00$ &$0.06 \pm 0.00$ & $1.12 \pm 0.01$ &$0.11 \pm 0.01$ & $0.48 \pm 0.02$ & 0.990 & $3.36 \pm 0.12$ & -1.13 & Y \\
 & CaT & 202 & $0.86 \pm 0.02$ &$0.05 \pm 0.01$ & $1.08 \pm 0.03$ &$0.11 \pm 0.01$ & $0.34 \pm 0.11$ & 0.999 & $2.50 \pm 0.27$ & -0.97 & Y \\
\hline
NGC 2768 & All & 195 & $0.70 \pm 0.04$ &$0.05 \pm 0.02$ & $1.00 \pm 0.03$ &$0.16 \pm 0.01$ & $0.17 \pm 0.09$ & 0.999 & $2.51 \pm 0.23$ & -0.78 & Y \\
 & CaT & 49 & $0.97 \pm 0.08$ &$0.10 \pm 0.03$ & $1.29 \pm 0.14$ &$0.03 \pm 0.04$ & $0.96 \pm 0.36$ & 0.645 & $4.40 \pm 1.51$ & 0.37 & N \\
\hline
NGC 3115 & All & 180 & $0.77 \pm 0.01$ &$0.09 \pm 0.01$ & $1.07 \pm 0.01$ &$0.07 \pm 0.01$ & $0.54 \pm 0.04$ & 0.999 & $3.88 \pm 0.26$ & -1.24 & Y \\
 & CaT & 122 & $0.79 \pm 0.01$ &$0.06 \pm 0.01$ & $1.09 \pm 0.01$ &$0.07 \pm 0.01$ & $0.47 \pm 0.05$ & 0.999 & $4.74 \pm 0.39$ & -1.38 & Y \\
\hline
NGC 3377 & All & 71 & $0.75 \pm 0.02$ &$0.05 \pm 0.02$ & $0.94 \pm 0.03$ &$0.07 \pm 0.02$ & $0.52 \pm 0.13$ & 0.997 & $3.07 \pm 0.54$ & -1.02 & Y \\
 & CaT & 84 & $0.79 \pm 0.02$ &$0.08 \pm 0.01$ & $1.01 \pm 0.03$ &$0.06 \pm 0.01$ & $0.67 \pm 0.10$ & 0.994 & $3.32 \pm 0.46$ & -1.01 & Y \\
\hline
NGC 4278 & All & 700 & $0.82 \pm 0.01$ &$0.08 \pm 0.01$ & $1.10 \pm 0.01$ &$0.10 \pm 0.01$ & $0.63 \pm 0.03$ & 0.999 & $3.28 \pm 0.20$ & -0.82 & Y \\
 & CaT & 150 & $0.80 \pm 0.03$ &$0.05 \pm 0.02$ & $1.02 \pm 0.04$ &$0.11 \pm 0.02$ & $0.40 \pm 0.13$ & 0.999 & $2.56 \pm 0.49$ & -1.02 & Y \\
\hline
NGC 4365 & All & 2159 & $0.79 \pm 0.00$ &$0.06 \pm 0.00$ & $1.05 \pm 0.01$ &$0.12 \pm 0.00$ & $0.36 \pm 0.02$ & 0.990 & $2.82 \pm 0.10$ & -1.01 & Y \\
 & CaT & 131 & $0.80 \pm 0.01$ &$0.04 \pm 0.01$ & $1.05 \pm 0.01$ &$0.10 \pm 0.01$ & $0.28 \pm 0.06$ & 0.999 & $3.34 \pm 0.31$ & -1.08 & Y \\
\hline
NGC 4494 & All & 127 & $0.86 \pm 0.02$ &$0.09 \pm 0.01$ & $1.07 \pm 0.03$ &$0.03 \pm 0.01$ & $0.80 \pm 0.09$ & 0.999 & $3.07 \pm 0.32$ & -0.97 & Y \\
 & CaT & 53 & $0.86 \pm 0.02$ &$0.08 \pm 0.01$ & $1.06 \pm 0.02$ &$0.04 \pm 0.01$ & $0.65 \pm 0.11$ & 0.986 & $3.20 \pm 0.53$ & -1.14 & Y \\
\hline
NGC 5846 & All & 894 & $0.76 \pm 0.02$ &$0.07 \pm 0.01$ & $1.01 \pm 0.02$ &$0.14 \pm 0.01$ & $0.34 \pm 0.08$ & 0.999 & $2.29 \pm 0.16$ & -0.89 & Y \\
 & CaT & 54 & $0.89 \pm 0.09$ &$0.13 \pm 0.05$ & $1.05 \pm 0.13$ &$0.14 \pm 0.05$ & $0.50 \pm 0.30$ & 0.022 & $1.18 \pm 1.14$ & -0.24 & N \\
\hline
NGC 7457 & All & 46 & $0.85 \pm 0.04$ &$0.17 \pm 0.05$ & $1.43 \pm 0.18$ &$0.04 \pm 0.08$ & $0.98 \pm 0.22$ & 0.551 & $4.70 \pm 1.49$ & 0.44 & N \\
\hline
\end{tabular}

\medskip
\emph{Notes} Column (1): Galaxy name.
Column (2): Whether all GC candidates or just GCs with CaT measurements. 
Column (3): Number of GCs.
Column (4): Mean colour of the blue subpopulation.
Column (5): Dispersion of the blue subpopulation.
Column (6): Mean colour of the red subpopulation.
Column (7): Dispersion of the red subpopulation.
Column (8): Fraction of the clusters in the blue population.
Column (9): p-value that a bimodal fit is preferred over a unimodal fit.
Column (10): Separation of the GMM peaks normalised by their width.
Column (11): The kurtosis of the sample.
Column (12): Bimodality of the sample. A p-value greater than 0.9, a separation of $D > 2$ and a negative kurtosis are all required for a sample to be considered bimodal.
\end{center}
\end{table*}

\section*{Acknowledgments}
We thank Karl Glazebrook, Lee Spitler and Albin Gaignette for providing data in advance of publication.
We thank Michael Murphy, Vincenzo Pota and Christina Blom for valuable discussions and comments.
We thank Mark Norris for assistance with the NGC 3115 Lick metallicities.
We thank the anonymous referee for proving useful comments.
This publication makes use of data products from the Two Micron All Sky Survey, which is a joint project of the University of Massachusetts and the Infrared Processing and Analysis Center/California Institute of Technology, funded by the National Aeronautics and Space Administration and the National Science Foundation.
This research has made use of the NASA/IPAC Extragalactic Database (NED) which is operated by the Jet Propulsion Laboratory, California Institute of Technology, under contract with the National Aeronautics and Space Administration. 
The analysis pipeline used to reduce the DEIMOS data was developed at UC Berkeley with support from NSF grant AST-0071048.
This publication made use of \textsc{PyRAF} and \textsc{PyFITS} which are products of the Space Telescope Science Institute, which is operated by AURA for NASA.
This research made use of \textsc{TOPCAT} \citep{2005ASPC..347...29T}.
This research is partially based on data from the MILES project.
Some of the data presented herein were obtained at the W. M. Keck Observatory, operated as a scientific partnership among the California Institute of Technology, the University of California and the National Aeronautics and Space Administration, and made possible by the generous financial support of the W. M. Keck Foundation.
The authors wish to recognize and acknowledge the very significant cultural role and reverence that the summit of Mauna Kea has always had within the indigenous Hawaiian community.  We are most fortunate to have the opportunity to conduct observations from this mountain. 
This research is based in part on data collected at Subaru Telescope, which is operated by the National Astronomical Observatory of Japan. 
This research is based on observations made with the NASA/ESA Hubble Space Telescope, obtained from the data archive at the Space Telescope Science Institute. STScI is operated by the Association of Universities for Research in Astronomy, Inc. under NASA contract NAS 5-26555.
CF acknowledges co-funding under the Marie Curie Actions of the European Commission (FP7-COFUND).
This material is based upon work supported by the National Science Foundation under grants AST-0808099 and AST-0909237.

\bibliographystyle{mn2e}
\bibliography{metal}

\begin{thebibliography}{116}
\expandafter\ifx\csname natexlab\endcsname\relax\def\natexlab#1{#1}\fi

\bibitem[{{Alves-Brito} {et~al.}(2011){Alves-Brito}, {Hau}, {Forbes},
  {Spitler}, {Strader}, {Brodie}, \& {Rhode}}]{2011MNRAS.417.1823A}
{Alves-Brito} A., {Hau} G.~K.~T., {Forbes} D.~A., {Spitler} L.~R., {Strader}
  J., {Brodie} J.~P., {Rhode} K.~L., 2011, \mnras, 417, 1823

\bibitem[{{Armandroff} \& {Zinn}(1988)}]{1988AJ.....96...92A}
{Armandroff} T.~E., {Zinn} R., 1988, \aj, 96, 92

\bibitem[{{Arnold} {et~al.}(2011){Arnold}, {Romanowsky}, {Brodie}, {Chomiuk},
  {Spitler}, {Strader}, {Benson}, \& {Forbes}}]{2011ApJ...736L..26A}
{Arnold} J.~A., {Romanowsky} A.~J., {Brodie} J.~P., {Chomiuk} L., {Spitler}
  L.~R., {Strader} J., {Benson} A.~J., {Forbes} D.~A., 2011, \apjl, 736, L26

\bibitem[{{Ashman} {et~al.}(1994){Ashman}, {Bird}, \&
  {Zepf}}]{1994AJ....108.2348A}
{Ashman} K.~M., {Bird} C.~M., {Zepf} S.~E., 1994, \aj, 108, 2348

\bibitem[{{Bailin} \& {Harris}(2009)}]{2009ApJ...695.1082B}
{Bailin} J., {Harris} W.~E., 2009, \apj, 695, 1082

\bibitem[{{Barmby} {et~al.}(2000){Barmby}, {Huchra}, {Brodie}, {Forbes},
  {Schroder}, \& {Grillmair}}]{2000AJ....119..727B}
{Barmby} P., {Huchra} J.~P., {Brodie} J.~P., {Forbes} D.~A., {Schroder} L.~L.,
  {Grillmair} C.~J., 2000, \aj, 119, 727

\bibitem[{{Bassino} {et~al.}(2006){Bassino}, {Richtler}, \&
  {Dirsch}}]{2006MNRAS.367..156B}
{Bassino} L.~P., {Richtler} T., {Dirsch} B., 2006, \mnras, 367, 156

\bibitem[{{Battaglia} {et~al.}(2008){Battaglia}, {Irwin}, {Tolstoy}, {Hill},
  {Helmi}, {Letarte}, \& {Jablonka}}]{2008MNRAS.383..183B}
{Battaglia} G., {Irwin} M., {Tolstoy} E., {Hill} V., {Helmi} A., {Letarte} B.,
  {Jablonka} P., 2008, \mnras, 383, 183

\bibitem[{{Beasley} {et~al.}(2008){Beasley}, {Bridges}, {Peng}, {Harris},
  {Harris}, {Forbes}, \& {Mackie}}]{2008MNRAS.386.1443B}
{Beasley} M.~A., {Bridges} T., {Peng} E., {Harris} W.~E., {Harris} G.~L.~H.,
  {Forbes} D.~A., {Mackie} G., 2008, \mnras, 386, 1443

\bibitem[{{Blakeslee} {et~al.}(2010){Blakeslee}, {Cantiello}, \&
  {Peng}}]{2010ApJ...710...51B}
{Blakeslee} J.~P., {Cantiello} M., {Peng} E.~W., 2010, \apj, 710, 51

\bibitem[{{Blakeslee} {et~al.}(2012){Blakeslee}, {Cho}, {Peng}, {Ferrarese},
  {Jord{\'a}n}, \& {Martel}}]{2012ApJ...746...88B}
{Blakeslee} J.~P., {Cho} H., {Peng} E.~W., {Ferrarese} L., {Jord{\'a}n} A.,
  {Martel} A.~R., 2012, \apj, 746, 88

\bibitem[{{Blom} {et~al.}(2012{\natexlab{a}}){Blom}, {Forbes}, {Brodie},
  {Foster}, {Romanowsky}, {Spitler}, \& {Strader}}]{Christina2012}
{Blom} C., {Forbes} D.~A., {Brodie} J.~P., {Foster} C., {Romanowsky} A.~J.,
  {Spitler} L., {Strader} J., 2012{\natexlab{a}}, {MNRAS}, in press

\bibitem[{{Blom} {et~al.}(2012{\natexlab{b}}){Blom}, {Spitler}, \&
  {Forbes}}]{2012MNRAS.420...37B}
{Blom} C., {Spitler} L.~R., {Forbes} D.~A., 2012{\natexlab{b}}, \mnras, 420, 37

\bibitem[{{Brodie} \& {Strader}(2006)}]{2006ARA&A..44..193B}
{Brodie} J.~P., {Strader} J., 2006, \araa, 44, 193

\bibitem[{{Brodie} {et~al.}(2005){Brodie}, {Strader}, {Denicol{\'o}},
  {Beasley}, {Cenarro}, {Larsen}, {Kuntschner}, \&
  {Forbes}}]{2005AJ....129.2643B}
{Brodie} J.~P., {Strader} J., {Denicol{\'o}} G., {Beasley} M.~A., {Cenarro}
  A.~J., {Larsen} S.~S., {Kuntschner} H., {Forbes} D.~A., 2005, \aj, 129, 2643

\bibitem[{{Brodie} {et~al.}(2012){Brodie}, {Usher}, {Conroy}, {Strader},
  {Arnold}, {Forbes}, \& {Romanowsky}}]{Jean2012}
{Brodie} J.~P., {Usher} C., {Conroy} C., {Strader} J., {Arnold} J.~A., {Forbes}
  D.~A., {Romanowsky} A.~J., 2012, {ApJL}, submitted

\bibitem[{{Caldwell} {et~al.}(2003){Caldwell}, {Rose}, \&
  {Concannon}}]{2003AJ....125.2891C}
{Caldwell} N., {Rose} J.~A., {Concannon} K.~D., 2003, \aj, 125, 2891

\bibitem[{{Caldwell} {et~al.}(2011){Caldwell}, {Schiavon}, {Morrison}, {Rose},
  \& {Harding}}]{2011AJ....141...61C}
{Caldwell} N., {Schiavon} R., {Morrison} H., {Rose} J.~A., {Harding} P., 2011,
  \aj, 141, 61

\bibitem[{{Cantiello} \& {Blakeslee}(2007)}]{2007ApJ...669..982C}
{Cantiello} M., {Blakeslee} J.~P., 2007, \apj, 669, 982

\bibitem[{{Cappellari} \& {Emsellem}(2004)}]{2004PASP..116..138C}
{Cappellari} M., {Emsellem} E., 2004, \pasp, 116, 138

\bibitem[{{Cappellari} {et~al.}(2011){Cappellari}, {Emsellem}, {Krajnovi{\'c}},
  {McDermid}, {Scott}, {Verdoes Kleijn}, {Young}, {Alatalo}, {Bacon}, {Blitz},
  {Bois}, {Bournaud}, {Bureau}, {Davies}, {Davis}, {de Zeeuw}, {Duc},
  {Khochfar}, {Kuntschner}, {Lablanche}, {Morganti}, {Naab}, {Oosterloo},
  {Sarzi}, {Serra}, \& {Weijmans}}]{2011MNRAS.413..813C}
{Cappellari} M., {Emsellem} E., {Krajnovi{\'c}} D., {McDermid} R.~M., {Scott}
  N., {Verdoes Kleijn} G.~A., {Young} L.~M., {Alatalo} K., {Bacon} R., {Blitz}
  L., {Bois} M., {Bournaud} F., {Bureau} M., {Davies} R.~L., {Davis} T.~A., {de
  Zeeuw} P.~T., {Duc} P.-A., {Khochfar} S., {Kuntschner} H., {Lablanche} P.-Y.,
  {Morganti} R., {Naab} T., {Oosterloo} T., {Sarzi} M., {Serra} P., {Weijmans}
  A.-M., 2011, \mnras, 413, 813

\bibitem[{{Cappellari} {et~al.}(2012){Cappellari}, {McDermid}, {Alatalo},
  {Blitz}, {Bois}, {Bournaud}, {Bureau}, {Crocker}, {Davies}, {Davis}, {de
  Zeeuw}, {Duc}, {Emsellem}, {Khochfar}, {Krajnovi{\'c}}, {Kuntschner},
  {Lablanche}, {Morganti}, {Naab}, {Oosterloo}, {Sarzi}, {Scott}, {Serra},
  {Weijmans}, \& {Young}}]{2012Natur.484..485C}
{Cappellari} M., {McDermid} R.~M., {Alatalo} K., {Blitz} L., {Bois} M.,
  {Bournaud} F., {Bureau} M., {Crocker} A.~F., {Davies} R.~L., {Davis} T.~A.,
  {de Zeeuw} P.~T., {Duc} P.-A., {Emsellem} E., {Khochfar} S., {Krajnovi{\'c}}
  D., {Kuntschner} H., {Lablanche} P.-Y., {Morganti} R., {Naab} T., {Oosterloo}
  T., {Sarzi} M., {Scott} N., {Serra} P., {Weijmans} A.-M., {Young} L.~M.,
  2012, \nat, 484, 485

\bibitem[{{Carretta} {et~al.}(2009){Carretta}, {Bragaglia}, {Gratton},
  {D'Orazi}, \& {Lucatello}}]{2009A&A...508..695C}
{Carretta} E., {Bragaglia} A., {Gratton} R., {D'Orazi} V., {Lucatello} S.,
  2009, \aap, 508, 695

\bibitem[{{Carretta} \& {Gratton}(1997)}]{1997A&AS..121...95C}
{Carretta} E., {Gratton} R.~G., 1997, \aaps, 121, 95

\bibitem[{{Cenarro} {et~al.}(2007){Cenarro}, {Beasley}, {Strader}, {Brodie}, \&
  {Forbes}}]{2007AJ....134..391C}
{Cenarro} A.~J., {Beasley} M.~A., {Strader} J., {Brodie} J.~P., {Forbes} D.~A.,
  2007, \aj, 134, 391

\bibitem[{{Cenarro} {et~al.}(2001){Cenarro}, {Cardiel}, {Gorgas}, {Peletier},
  {Vazdekis}, \& {Prada}}]{2001MNRAS.326..959C}
{Cenarro} A.~J., {Cardiel} N., {Gorgas} J., {Peletier} R.~F., {Vazdekis} A.,
  {Prada} F., 2001, \mnras, 326, 959

\bibitem[{{Cenarro} {et~al.}(2002){Cenarro}, {Gorgas}, {Cardiel}, {Vazdekis},
  \& {Peletier}}]{2002MNRAS.329..863C}
{Cenarro} A.~J., {Gorgas} J., {Cardiel} N., {Vazdekis} A., {Peletier} R.~F.,
  2002, \mnras, 329, 863

\bibitem[{{Chies-Santos} {et~al.}(2012){Chies-Santos}, {Larsen}, {Cantiello},
  {Strader}, {Kuntschner}, {Wehner}, \& {Brodie}}]{2012A&A...539A..54C}
{Chies-Santos} A.~L., {Larsen} S.~S., {Cantiello} M., {Strader} J.,
  {Kuntschner} H., {Wehner} E.~M., {Brodie} J.~P., 2012, \aap, 539, A54

\bibitem[{{Chomiuk} {et~al.}(2008){Chomiuk}, {Strader}, \&
  {Brodie}}]{2008AJ....136..234C}
{Chomiuk} L., {Strader} J., {Brodie} J.~P., 2008, \aj, 136, 234

\bibitem[{{Cohen} {et~al.}(2003){Cohen}, {Blakeslee}, \&
  {C{\^o}t{\'e}}}]{2003ApJ...592..866C}
{Cohen} J.~G., {Blakeslee} J.~P., {C{\^o}t{\'e}} P., 2003, \apj, 592, 866

\bibitem[{{Cohen} {et~al.}(1998){Cohen}, {Blakeslee}, \&
  {Ryzhov}}]{1998ApJ...496..808C}
{Cohen} J.~G., {Blakeslee} J.~P., {Ryzhov} A., 1998, \apj, 496, 808

\bibitem[{{Conroy} {et~al.}(2009){Conroy}, {Gunn}, \&
  {White}}]{2009ApJ...699..486C}
{Conroy} C., {Gunn} J.~E., {White} M., 2009, \apj, 699, 486

\bibitem[{{Cooper} {et~al.}(2012){Cooper}, {Newman}, {Davis}, {Finkbeiner}, \&
  {Gerke}}]{2012ascl.soft03003C}
{Cooper} M.~C., {Newman} J.~A., {Davis} M., {Finkbeiner} D.~P., {Gerke} B.~F.,
  2012, in Astrophysics Source Code Library, record ascl:1203.003, p. 3003

\bibitem[{{C{\^o}t{\'e}} {et~al.}(2003){C{\^o}t{\'e}}, {McLaughlin}, {Cohen},
  \& {Blakeslee}}]{2003ApJ...591..850C}
{C{\^o}t{\'e}} P., {McLaughlin} D.~E., {Cohen} J.~G., {Blakeslee} J.~P., 2003,
  \apj, 591, 850

\bibitem[{{Dotter} {et~al.}(2011){Dotter}, {Sarajedini}, \&
  {Anderson}}]{2011ApJ...738...74D}
{Dotter} A., {Sarajedini} A., {Anderson} J., 2011, \apj, 738, 74

\bibitem[{{Faber} {et~al.}(2003){Faber}, {Phillips}, {Kibrick}, {Alcott},
  {Allen}, {Burrous}, {Cantrall}, {Clarke}, {Coil}, {Cowley}, {Davis}, {Deich},
  {Dietsch}, {Gilmore}, {Harper}, {Hilyard}, {Lewis}, {McVeigh}, {Newman},
  {Osborne}, {Schiavon}, {Stover}, {Tucker}, {Wallace}, {Wei}, {Wirth}, \&
  {Wright}}]{2003SPIE.4841.1657F}
{Faber} S.~M., {Phillips} A.~C., {Kibrick} R.~I., {Alcott} B., {Allen} S.~L.,
  {Burrous} J., {Cantrall} T., {Clarke} D., {Coil} A.~L., {Cowley} D.~J.,
  {Davis} M., {Deich} W.~T.~S., {Dietsch} K., {Gilmore} D.~K., {Harper} C.~A.,
  {Hilyard} D.~F., {Lewis} J.~P., {McVeigh} M., {Newman} J., {Osborne} J.,
  {Schiavon} R., {Stover} R.~J., {Tucker} D., {Wallace} V., {Wei} M., {Wirth}
  G., {Wright} C.~A., 2003, in Presented at the Society of Photo-Optical
  Instrumentation Engineers (SPIE) Conference, Vol. 4841, Society of
  Photo-Optical Instrumentation Engineers (SPIE) Conference Series, {M.~Iye \&
  A.~F.~M.~Moorwood}, ed., pp. 1657--1669

\bibitem[{{Faifer} {et~al.}(2011){Faifer}, {Forte}, {Norris}, {Bridges},
  {Forbes}, {Zepf}, {Beasley}, {Gebhardt}, {Hanes}, \&
  {Sharples}}]{2011MNRAS.416..155F}
{Faifer} F.~R., {Forte} J.~C., {Norris} M.~A., {Bridges} T., {Forbes} D.~A.,
  {Zepf} S.~E., {Beasley} M., {Gebhardt} K., {Hanes} D.~A., {Sharples} R.~M.,
  2011, \mnras, 416, 155

\bibitem[{{Forbes} {et~al.}(2001){Forbes}, {Beasley}, {Brodie}, \&
  {Kissler-Patig}}]{2001ApJ...563L.143F}
{Forbes} D.~A., {Beasley} M.~A., {Brodie} J.~P., {Kissler-Patig} M., 2001,
  \apjl, 563, L143

\bibitem[{{Forbes} \& {Bridges}(2010)}]{2010MNRAS.404.1203F}
{Forbes} D.~A., {Bridges} T., 2010, \mnras, 404, 1203

\bibitem[{{Forbes} {et~al.}(1996){Forbes}, {Brodie}, \&
  {Huchra}}]{1996AJ....112.2448F}
{Forbes} D.~A., {Brodie} J.~P., {Huchra} J., 1996, \aj, 112, 2448

\bibitem[{{Forbes} {et~al.}(2012){Forbes}, {Ponman}, \&
  {O'Sullivan}}]{Duncan2012}
{Forbes} D.~A., {Ponman} T., {O'Sullivan} E., 2012, {MNRAS}, in press
  (arXiv:1205.5315)

\bibitem[{{Forbes} {et~al.}(2006){Forbes}, {S{\'a}nchez-Bl{\'a}zquez}, {Phan},
  {Brodie}, {Strader}, \& {Spitler}}]{2006MNRAS.366.1230F}
{Forbes} D.~A., {S{\'a}nchez-Bl{\'a}zquez} P., {Phan} A.~T.~T., {Brodie} J.~P.,
  {Strader} J., {Spitler} L., 2006, \mnras, 366, 1230

\bibitem[{{Foster} {et~al.}(2010){Foster}, {Forbes}, {Proctor}, {Strader},
  {Brodie}, \& {Spitler}}]{2010AJ....139.1566F}
{Foster} C., {Forbes} D.~A., {Proctor} R.~N., {Strader} J., {Brodie} J.~P.,
  {Spitler} L.~R., 2010, \aj, 139, 1566

\bibitem[{{Foster} {et~al.}(2011){Foster}, {Spitler}, {Romanowsky}, {Forbes},
  {Pota}, {Bekki}, {Strader}, {Proctor}, {Arnold}, \&
  {Brodie}}]{2011MNRAS.415.3393F}
{Foster} C., {Spitler} L.~R., {Romanowsky} A.~J., {Forbes} D.~A., {Pota} V.,
  {Bekki} K., {Strader} J., {Proctor} R.~N., {Arnold} J.~A., {Brodie} J.~P.,
  2011, \mnras, 415, 3393

\bibitem[{{Galleti} {et~al.}(2009){Galleti}, {Bellazzini}, {Buzzoni},
  {Federici}, \& {Fusi Pecci}}]{2009A&A...508.1285G}
{Galleti} S., {Bellazzini} M., {Buzzoni} A., {Federici} L., {Fusi Pecci} F.,
  2009, \aap, 508, 1285

\bibitem[{{Galleti} {et~al.}(2006){Galleti}, {Federici}, {Bellazzini},
  {Buzzoni}, \& {Pecci}}]{2006ApJ...650L.107G}
{Galleti} S., {Federici} L., {Bellazzini} M., {Buzzoni} A., {Pecci} F.~F.,
  2006, \apjl, 650, L107

\bibitem[{{Geisler} {et~al.}(1996){Geisler}, {Lee}, \&
  {Kim}}]{1996AJ....111.1529G}
{Geisler} D., {Lee} M.~G., {Kim} E., 1996, \aj, 111, 1529

\bibitem[{{Girardi} {et~al.}(2000){Girardi}, {Bressan}, {Bertelli}, \&
  {Chiosi}}]{2000A&AS..141..371G}
{Girardi} L., {Bressan} A., {Bertelli} G., {Chiosi} C., 2000, \aaps, 141, 371

\bibitem[{{Hargis} {et~al.}(2011){Hargis}, {Rhode}, {Strader}, \&
  {Brodie}}]{2011ApJ...738..113H}
{Hargis} J.~R., {Rhode} K.~L., {Strader} J., {Brodie} J.~P., 2011, \apj, 738,
  113

\bibitem[{{Harris}(1996)}]{1996AJ....112.1487H}
{Harris} W.~E., 1996, \aj, 112, 1487

\bibitem[{{Harris}(2009)}]{2009ApJ...703..939H}
---, 2009, \apj, 703, 939

\bibitem[{{Harris}(2010)}]{harrisMWcatalogue}
---, 2010, {preprint (arXiv:1012.3224)}

\bibitem[{{Harris} \& {Canterna}(1979)}]{1979ApJ...231L..19H}
{Harris} W.~E., {Canterna} R., 1979, \apjl, 231, L19

\bibitem[{{Harris} {et~al.}(2006){Harris}, {Whitmore}, {Karakla}, {Oko{\'n}},
  {Baum}, {Hanes}, \& {Kavelaars}}]{2006ApJ...636...90H}
{Harris} W.~E., {Whitmore} B.~C., {Karakla} D., {Oko{\'n}} W., {Baum} W.~A.,
  {Hanes} D.~A., {Kavelaars} J.~J., 2006, \apj, 636, 90

\bibitem[{{Jordi} {et~al.}(2006){Jordi}, {Grebel}, \&
  {Ammon}}]{2006A&A...460..339J}
{Jordi} K., {Grebel} E.~K., {Ammon} K., 2006, \aap, 460, 339

\bibitem[{{Kroupa}(2001)}]{2001MNRAS.322..231K}
{Kroupa} P., 2001, \mnras, 322, 231

\bibitem[{{Kundu} \& {Whitmore}(2001)}]{2001AJ....121.2950K}
{Kundu} A., {Whitmore} B.~C., 2001, \aj, 121, 2950

\bibitem[{{Kundu} \& {Zepf}(2007)}]{2007ApJ...660L.109K}
{Kundu} A., {Zepf} S.~E., 2007, \apjl, 660, L109

\bibitem[{{Kuntschner} {et~al.}(2002){Kuntschner}, {Ziegler}, {Sharples},
  {Worthey}, \& {Fricke}}]{2002A&A...395..761K}
{Kuntschner} H., {Ziegler} B.~L., {Sharples} R.~M., {Worthey} G., {Fricke}
  K.~J., 2002, \aap, 395, 761

\bibitem[{{Larsen} {et~al.}(2001){Larsen}, {Brodie}, {Huchra}, {Forbes}, \&
  {Grillmair}}]{2001AJ....121.2974L}
{Larsen} S.~S., {Brodie} J.~P., {Huchra} J.~P., {Forbes} D.~A., {Grillmair}
  C.~J., 2001, \aj, 121, 2974

\bibitem[{{Lee} {et~al.}(2010){Lee}, {Park}, {Hwang}, {Arimoto}, {Tamura}, \&
  {Onodera}}]{2010ApJ...709.1083L}
{Lee} M.~G., {Park} H.~S., {Hwang} H.~S., {Arimoto} N., {Tamura} N., {Onodera}
  M., 2010, \apj, 709, 1083

\bibitem[{{Mackey} {et~al.}(2007){Mackey}, {Huxor}, {Ferguson}, {Tanvir},
  {Irwin}, {Ibata}, {Bridges}, {Johnson}, \& {Lewis}}]{2007ApJ...655L..85M}
{Mackey} A.~D., {Huxor} A., {Ferguson} A.~M.~N., {Tanvir} N.~R., {Irwin} M.,
  {Ibata} R., {Bridges} T., {Johnson} R.~A., {Lewis} G., 2007, \apjl, 655, L85

\bibitem[{{Maraston}(2005)}]{2005MNRAS.362..799M}
{Maraston} C., 2005, \mnras, 362, 799

\bibitem[{{Mei} {et~al.}(2007){Mei}, {Blakeslee}, {C{\^o}t{\'e}}, {Tonry},
  {West}, {Ferrarese}, {Jord{\'a}n}, {Peng}, {Anthony}, \&
  {Merritt}}]{2007ApJ...655..144M}
{Mei} S., {Blakeslee} J.~P., {C{\^o}t{\'e}} P., {Tonry} J.~L., {West} M.~J.,
  {Ferrarese} L., {Jord{\'a}n} A., {Peng} E.~W., {Anthony} A., {Merritt} D.,
  2007, \apj, 655, 144

\bibitem[{{Mendel} {et~al.}(2007){Mendel}, {Proctor}, \&
  {Forbes}}]{2007MNRAS.379.1618M}
{Mendel} J.~T., {Proctor} R.~N., {Forbes} D.~A., 2007, \mnras, 379, 1618

\bibitem[{{Merrett} {et~al.}(2003){Merrett}, {Kuijken}, {Merrifield},
  {Romanowsky}, {Douglas}, {Napolitano}, {Arnaboldi}, {Capaccioli}, {Freeman},
  {Gerhard}, {Evans}, {Wilkinson}, {Halliday}, {Bridges}, \&
  {Carter}}]{2003MNRAS.346L..62M}
{Merrett} H.~R., {Kuijken} K., {Merrifield} M.~R., {Romanowsky} A.~J.,
  {Douglas} N.~G., {Napolitano} N.~R., {Arnaboldi} M., {Capaccioli} M.,
  {Freeman} K.~C., {Gerhard} O., {Evans} N.~W., {Wilkinson} M.~I., {Halliday}
  C., {Bridges} T.~J., {Carter} D., 2003, \mnras, 346, L62

\bibitem[{{Mieske} {et~al.}(2006){Mieske}, {Jord{\'a}n}, {C{\^o}t{\'e}},
  {Kissler-Patig}, {Peng}, {Ferrarese}, {Blakeslee}, {Mei}, {Merritt}, {Tonry},
  \& {West}}]{2006ApJ...653..193M}
{Mieske} S., {Jord{\'a}n} A., {C{\^o}t{\'e}} P., {Kissler-Patig} M., {Peng}
  E.~W., {Ferrarese} L., {Blakeslee} J.~P., {Mei} S., {Merritt} D., {Tonry}
  J.~L., {West} M.~J., 2006, \apj, 653, 193

\bibitem[{{Miyazaki} {et~al.}(2002){Miyazaki}, {Komiyama}, {Sekiguchi},
  {Okamura}, {Doi}, {Furusawa}, {Hamabe}, {Imi}, {Kimura}, {Nakata}, {Okada},
  {Ouchi}, {Shimasaku}, {Yagi}, \& {Yasuda}}]{2002PASJ...54..833M}
{Miyazaki} S., {Komiyama} Y., {Sekiguchi} M., {Okamura} S., {Doi} M.,
  {Furusawa} H., {Hamabe} M., {Imi} K., {Kimura} M., {Nakata} F., {Okada} N.,
  {Ouchi} M., {Shimasaku} K., {Yagi} M., {Yasuda} N., 2002, \pasj, 54, 833

\bibitem[{{Muratov} \& {Gnedin}(2010)}]{2010ApJ...718.1266M}
{Muratov} A.~L., {Gnedin} O.~Y., 2010, \apj, 718, 1266

\bibitem[{{Nelan} {et~al.}(2005){Nelan}, {Smith}, {Hudson}, {Wegner}, {Lucey},
  {Moore}, {Quinney}, \& {Suntzeff}}]{2005ApJ...632..137N}
{Nelan} J.~E., {Smith} R.~J., {Hudson} M.~J., {Wegner} G.~A., {Lucey} J.~R.,
  {Moore} S.~A.~W., {Quinney} S.~J., {Suntzeff} N.~B., 2005, \apj, 632, 137

\bibitem[{{Newman} {et~al.}(2012){Newman}, {Cooper}, {Davis}, {Faber}, {Coil},
  {Guhathakurta}, {Koo}, {Phillips}, {Conroy}, {Dutton}, {Finkbeiner}, {Gerke},
  {Rosario}, {Weiner}, {Willmer}, {Yan}, {Harker}, {Kassin}, {Konidaris},
  {Lai}, {Madgwick}, {Noeske}, {Wirth}, {Connolly}, {Kaiser}, {Kirby},
  {Lemaux}, {Lin}, {Lotz}, {Luppino}, {Marinoni}, {Matthews}, {Metevier}, \&
  {Schiavon}}]{Newman2012}
{Newman} J.~A., {Cooper} M.~C., {Davis} M., {Faber} S.~M., {Coil} A.~L.,
  {Guhathakurta} P., {Koo} D.~C., {Phillips} A.~C., {Conroy} C., {Dutton}
  A.~A., {Finkbeiner} D.~P., {Gerke} B.~F., {Rosario} D.~J., {Weiner} B.~J.,
  {Willmer} C.~N.~A., {Yan} R., {Harker} J.~J., {Kassin} S.~A., {Konidaris}
  N.~P., {Lai} K., {Madgwick} D.~S., {Noeske} K.~G., {Wirth} G.~D., {Connolly}
  A.~J., {Kaiser} N., {Kirby} E.~N., {Lemaux} B.~C., {Lin} L., {Lotz} J.~M.,
  {Luppino} G.~A., {Marinoni} C., {Matthews} D.~J., {Metevier} A., {Schiavon}
  R.~P., 2012, {ApJS}, submitted (arXiv:1203.3192)

\bibitem[{{Oke} {et~al.}(1995){Oke}, {Cohen}, {Carr}, {Cromer}, {Dingizian},
  {Harris}, {Labrecque}, {Lucinio}, {Schaal}, {Epps}, \&
  {Miller}}]{1995PASP..107..375O}
{Oke} J.~B., {Cohen} J.~G., {Carr} M., {Cromer} J., {Dingizian} A., {Harris}
  F.~H., {Labrecque} S., {Lucinio} R., {Schaal} W., {Epps} H., {Miller} J.,
  1995, \pasp, 107, 375

\bibitem[{{Peacock} {et~al.}(2010){Peacock}, {Maccarone}, {Knigge}, {Kundu},
  {Waters}, {Zepf}, \& {Zurek}}]{2010MNRAS.402..803P}
{Peacock} M.~B., {Maccarone} T.~J., {Knigge} C., {Kundu} A., {Waters} C.~Z.,
  {Zepf} S.~E., {Zurek} D.~R., 2010, \mnras, 402, 803

\bibitem[{{Peng} {et~al.}(2006){Peng}, {Jord{\'a}n}, {C{\^o}t{\'e}},
  {Blakeslee}, {Ferrarese}, {Mei}, {West}, {Merritt}, {Milosavljevi{\'c}}, \&
  {Tonry}}]{2006ApJ...639...95P}
{Peng} E.~W., {Jord{\'a}n} A., {C{\^o}t{\'e}} P., {Blakeslee} J.~P.,
  {Ferrarese} L., {Mei} S., {West} M.~J., {Merritt} D., {Milosavljevi{\'c}} M.,
  {Tonry} J.~L., 2006, \apj, 639, 95

\bibitem[{{Perina} {et~al.}(2009){Perina}, {Federici}, {Bellazzini},
  {Cacciari}, {Fusi Pecci}, \& {Galleti}}]{2009A&A...507.1375P}
{Perina} S., {Federici} L., {Bellazzini} M., {Cacciari} C., {Fusi Pecci} F.,
  {Galleti} S., 2009, \aap, 507, 1375

\bibitem[{{Perina} {et~al.}(2011){Perina}, {Galleti}, {Fusi Pecci},
  {Bellazzini}, {Federici}, \& {Buzzoni}}]{2011A&A...531A.155P}
{Perina} S., {Galleti} S., {Fusi Pecci} F., {Bellazzini} M., {Federici} L.,
  {Buzzoni} A., 2011, \aap, 531, A155

\bibitem[{{Pota} {et~al.}(2012){Pota}, {Forbes}, {Romanowsky}, {Brodie},
  {Spitler}, {Strader}, {Foster}, {Arnold}, {Benson}, {Blom}, {Hargis}, \&
  {Usher}}]{Vince2012}
{Pota} V., {Forbes} D.~A., {Romanowsky} A.~J., {Brodie} J.~P., {Spitler} L.,
  {Strader} J., {Foster} C., {Arnold} J.~A., {Benson} A., {Blom} C., {Hargis}
  J.~R., {Usher} C., 2012, {MNRAS}, submitted

\bibitem[{{Puzia} {et~al.}(2005){Puzia}, {Kissler-Patig}, {Thomas}, {Maraston},
  {Saglia}, {Bender}, {Goudfrooij}, \& {Hempel}}]{2005A&A...439..997P}
{Puzia} T.~H., {Kissler-Patig} M., {Thomas} D., {Maraston} C., {Saglia} R.~P.,
  {Bender} R., {Goudfrooij} P., {Hempel} M., 2005, \aap, 439, 997

\bibitem[{{Raimondo} {et~al.}(2005){Raimondo}, {Brocato}, {Cantiello}, \&
  {Capaccioli}}]{2005AJ....130.2625R}
{Raimondo} G., {Brocato} E., {Cantiello} M., {Capaccioli} M., 2005, \aj, 130,
  2625

\bibitem[{{Rich} {et~al.}(2005){Rich}, {Corsi}, {Cacciari}, {Federici}, {Fusi
  Pecci}, {Djorgovski}, \& {Freedman}}]{2005AJ....129.2670R}
{Rich} R.~M., {Corsi} C.~E., {Cacciari} C., {Federici} L., {Fusi Pecci} F.,
  {Djorgovski} S.~G., {Freedman} W.~L., 2005, \aj, 129, 2670

\bibitem[{{Richtler}(2006)}]{2006BASI...34...83R}
{Richtler} T., 2006, Bulletin of the Astronomical Society of India, 34, 83

\bibitem[{{Romanowsky} {et~al.}(2009){Romanowsky}, {Strader}, {Spitler},
  {Johnson}, {Brodie}, {Forbes}, \& {Ponman}}]{2009AJ....137.4956R}
{Romanowsky} A.~J., {Strader} J., {Spitler} L.~R., {Johnson} R., {Brodie}
  J.~P., {Forbes} D.~A., {Ponman} T., 2009, \aj, 137, 4956

\bibitem[{{Rutledge} {et~al.}(1997){Rutledge}, {Hesser}, \&
  {Stetson}}]{1997PASP..109..907R}
{Rutledge} G.~A., {Hesser} J.~E., {Stetson} P.~B., 1997, \pasp, 109, 907

\bibitem[{{Salpeter}(1955)}]{1955ApJ...121..161S}
{Salpeter} E.~E., 1955, \apj, 121, 161

\bibitem[{{Scalo}(1998)}]{1998ASPC..142..201S}
{Scalo} J., 1998, in Astronomical Society of the Pacific Conference Series,
  Vol. 142, The Stellar Initial Mass Function (38th Herstmonceux Conference),
  {G.~Gilmore \& D.~Howell}, ed., p. 201

\bibitem[{{Schlegel} {et~al.}(1998){Schlegel}, {Finkbeiner}, \&
  {Davis}}]{1998ApJ...500..525S}
{Schlegel} D.~J., {Finkbeiner} D.~P., {Davis} M., 1998, \apj, 500, 525

\bibitem[{{Schuberth} {et~al.}(2010){Schuberth}, {Richtler}, {Hilker},
  {Dirsch}, {Bassino}, {Romanowsky}, \& {Infante}}]{2010A&A...513A..52S}
{Schuberth} Y., {Richtler} T., {Hilker} M., {Dirsch} B., {Bassino} L.~P.,
  {Romanowsky} A.~J., {Infante} L., 2010, \aap, 513, A52

\bibitem[{{Sinnott} {et~al.}(2010){Sinnott}, {Hou}, {Anderson}, {Harris}, \&
  {Woodley}}]{2010AJ....140.2101S}
{Sinnott} B., {Hou} A., {Anderson} R., {Harris} W.~E., {Woodley} K.~A., 2010,
  \aj, 140, 2101

\bibitem[{{Skrutskie} {et~al.}(2006){Skrutskie}, {Cutri}, {Stiening},
  {Weinberg}, {Schneider}, {Carpenter}, {Beichman}, {Capps}, {Chester},
  {Elias}, {Huchra}, {Liebert}, {Lonsdale}, {Monet}, {Price}, {Seitzer},
  {Jarrett}, {Kirkpatrick}, {Gizis}, {Howard}, {Evans}, {Fowler}, {Fullmer},
  {Hurt}, {Light}, {Kopan}, {Marsh}, {McCallon}, {Tam}, {Van Dyk}, \&
  {Wheelock}}]{2006AJ....131.1163S}
{Skrutskie} M.~F., {Cutri} R.~M., {Stiening} R., {Weinberg} M.~D., {Schneider}
  S., {Carpenter} J.~M., {Beichman} C., {Capps} R., {Chester} T., {Elias} J.,
  {Huchra} J., {Liebert} J., {Lonsdale} C., {Monet} D.~G., {Price} S.,
  {Seitzer} P., {Jarrett} T., {Kirkpatrick} J.~D., {Gizis} J.~E., {Howard} E.,
  {Evans} T., {Fowler} J., {Fullmer} L., {Hurt} R., {Light} R., {Kopan} E.~L.,
  {Marsh} K.~A., {McCallon} H.~L., {Tam} R., {Van Dyk} S., {Wheelock} S., 2006,
  \aj, 131, 1163

\bibitem[{{Smith} {et~al.}(2012){Smith}, {Lucey}, {Price}, {Hudson}, \&
  {Phillipps}}]{2012MNRAS.419.3167S}
{Smith} R.~J., {Lucey} J.~R., {Price} J., {Hudson} M.~J., {Phillipps} S., 2012,
  \mnras, 419, 3167

\bibitem[{{Spitler} {et~al.}(2008{\natexlab{a}}){Spitler}, {Forbes}, \&
  {Beasley}}]{2008MNRAS.389.1150S}
{Spitler} L.~R., {Forbes} D.~A., {Beasley} M.~A., 2008{\natexlab{a}}, \mnras,
  389, 1150

\bibitem[{{Spitler} {et~al.}(2008{\natexlab{b}}){Spitler}, {Forbes}, {Strader},
  {Brodie}, \& {Gallagher}}]{2008MNRAS.385..361S}
{Spitler} L.~R., {Forbes} D.~A., {Strader} J., {Brodie} J.~P., {Gallagher}
  J.~S., 2008{\natexlab{b}}, \mnras, 385, 361

\bibitem[{{Spitler} {et~al.}(2012){Spitler}, {Romanowsky}, {Diemand},
  {Strader}, {Forbes}, {Moore}, \& {Brodie}}]{2012MNRAS.423.2177S}
{Spitler} L.~R., {Romanowsky} A.~J., {Diemand} J., {Strader} J., {Forbes}
  D.~A., {Moore} B., {Brodie} J.~P., 2012, \mnras, 423, 2177

\bibitem[{{Strader} {et~al.}(2007){Strader}, {Beasley}, \&
  {Brodie}}]{2007AJ....133.2015S}
{Strader} J., {Beasley} M.~A., {Brodie} J.~P., 2007, \aj, 133, 2015

\bibitem[{{Strader} {et~al.}(2005){Strader}, {Brodie}, {Cenarro}, {Beasley}, \&
  {Forbes}}]{2005AJ....130.1315S}
{Strader} J., {Brodie} J.~P., {Cenarro} A.~J., {Beasley} M.~A., {Forbes} D.~A.,
  2005, \aj, 130, 1315

\bibitem[{{Strader} {et~al.}(2006){Strader}, {Brodie}, {Spitler}, \&
  {Beasley}}]{2006AJ....132.2333S}
{Strader} J., {Brodie} J.~P., {Spitler} L., {Beasley} M.~A., 2006, \aj, 132,
  2333

\bibitem[{{Strader} {et~al.}(2011){Strader}, {Romanowsky}, {Brodie}, {Spitler},
  {Beasley}, {Arnold}, {Tamura}, {Sharples}, \&
  {Arimoto}}]{2011ApJS..197...33S}
{Strader} J., {Romanowsky} A.~J., {Brodie} J.~P., {Spitler} L.~R., {Beasley}
  M.~A., {Arnold} J.~A., {Tamura} N., {Sharples} R.~M., {Arimoto} N., 2011,
  \apjs, 197, 33

\bibitem[{{Taylor}(2005)}]{2005ASPC..347...29T}
{Taylor} M.~B., 2005, in Astronomical Society of the Pacific Conference Series,
  Vol. 347, Astronomical Data Analysis Software and Systems XIV, {Shopbell} P.,
  {Britton} M., {Ebert} R., eds., p.~29

\bibitem[{{Terlevich} \& {Forbes}(2002)}]{2002MNRAS.330..547T}
{Terlevich} A.~I., {Forbes} D.~A., 2002, \mnras, 330, 547

\bibitem[{{Thomas} {et~al.}(2003){Thomas}, {Maraston}, \&
  {Bender}}]{2003MNRAS.339..897T}
{Thomas} D., {Maraston} C., {Bender} R., 2003, \mnras, 339, 897

\bibitem[{{Thomas} {et~al.}(2005){Thomas}, {Maraston}, {Bender}, \& {Mendes de
  Oliveira}}]{2005ApJ...621..673T}
{Thomas} D., {Maraston} C., {Bender} R., {Mendes de Oliveira} C., 2005, \apj,
  621, 673

\bibitem[{{Thomas} {et~al.}(2004){Thomas}, {Maraston}, \&
  {Korn}}]{2004MNRAS.351L..19T}
{Thomas} D., {Maraston} C., {Korn} A., 2004, \mnras, 351, L19

\bibitem[{{Tonry} {et~al.}(2001){Tonry}, {Dressler}, {Blakeslee}, {Ajhar},
  {Fletcher}, {Luppino}, {Metzger}, \& {Moore}}]{2001ApJ...546..681T}
{Tonry} J.~L., {Dressler} A., {Blakeslee} J.~P., {Ajhar} E.~A., {Fletcher}
  A.~B., {Luppino} G.~A., {Metzger} M.~R., {Moore} C.~B., 2001, \apj, 546, 681

\bibitem[{{Trager} {et~al.}(2000){Trager}, {Faber}, {Worthey}, \&
  {Gonz{\'a}lez}}]{2000AJ....120..165T}
{Trager} S.~C., {Faber} S.~M., {Worthey} G., {Gonz{\'a}lez} J.~J., 2000, \aj,
  120, 165

\bibitem[{{van Dokkum} \& {Conroy}(2010)}]{2010Natur.468..940V}
{van Dokkum} P.~G., {Conroy} C., 2010, \nat, 468, 940

\bibitem[{{Vazdekis} {et~al.}(2003){Vazdekis}, {Cenarro}, {Gorgas}, {Cardiel},
  \& {Peletier}}]{2003MNRAS.340.1317V}
{Vazdekis} A., {Cenarro} A.~J., {Gorgas} J., {Cardiel} N., {Peletier} R.~F.,
  2003, \mnras, 340, 1317

\bibitem[{{Vazdekis} {et~al.}(2010){Vazdekis}, {S{\'a}nchez-Bl{\'a}zquez},
  {Falc{\'o}n-Barroso}, {Cenarro}, {Beasley}, {Cardiel}, {Gorgas}, \&
  {Peletier}}]{2010MNRAS.404.1639V}
{Vazdekis} A., {S{\'a}nchez-Bl{\'a}zquez} P., {Falc{\'o}n-Barroso} J.,
  {Cenarro} A.~J., {Beasley} M.~A., {Cardiel} N., {Gorgas} J., {Peletier}
  R.~F., 2010, \mnras, 404, 1639

\bibitem[{{Villegas} {et~al.}(2010){Villegas}, {Jord{\'a}n}, {Peng},
  {Blakeslee}, {C{\^o}t{\'e}}, {Ferrarese}, {Kissler-Patig}, {Mei}, {Infante},
  {Tonry}, \& {West}}]{2010ApJ...717..603V}
{Villegas} D., {Jord{\'a}n} A., {Peng} E.~W., {Blakeslee} J.~P., {C{\^o}t{\'e}}
  P., {Ferrarese} L., {Kissler-Patig} M., {Mei} S., {Infante} L., {Tonry}
  J.~L., {West} M.~J., 2010, \apj, 717, 603

\bibitem[{{Woodley} {et~al.}(2010{\natexlab{a}}){Woodley}, {G{\'o}mez},
  {Harris}, {Geisler}, \& {Harris}}]{2010AJ....139.1871W}
{Woodley} K.~A., {G{\'o}mez} M., {Harris} W.~E., {Geisler} D., {Harris}
  G.~L.~H., 2010{\natexlab{a}}, \aj, 139, 1871

\bibitem[{{Woodley} {et~al.}(2010{\natexlab{b}}){Woodley}, {Harris}, {Puzia},
  {G{\'o}mez}, {Harris}, \& {Geisler}}]{2010ApJ...708.1335W}
{Woodley} K.~A., {Harris} W.~E., {Puzia} T.~H., {G{\'o}mez} M., {Harris}
  G.~L.~H., {Geisler} D., 2010{\natexlab{b}}, \apj, 708, 1335

\bibitem[{{Worthey}(1994)}]{1994ApJS...95..107W}
{Worthey} G., 1994, \apjs, 95, 107

\bibitem[{{Worthey} {et~al.}(1994){Worthey}, {Faber}, {Gonzalez}, \&
  {Burstein}}]{1994ApJS...94..687W}
{Worthey} G., {Faber} S.~M., {Gonzalez} J.~J., {Burstein} D., 1994, \apjs, 94,
  687

\bibitem[{{Yoon} {et~al.}(2006){Yoon}, {Yi}, \& {Lee}}]{2006Sci...311.1129Y}
{Yoon} S., {Yi} S.~K., {Lee} Y., 2006, Science, 311, 1129

\bibitem[{{Yoon} {et~al.}(2011){Yoon}, {Lee}, {Blakeslee}, {Peng}, {Sohn},
  {Cho}, {Kim}, {Chung}, {Kim}, \& {Lee}}]{2011ApJ...743..150Y}
{Yoon} S.-J., {Lee} S.-Y., {Blakeslee} J.~P., {Peng} E.~W., {Sohn} S.~T., {Cho}
  J., {Kim} H.-S., {Chung} C., {Kim} S., {Lee} Y.-W., 2011, \apj, 743, 150

\bibitem[{{Zinn}(1985)}]{1985ApJ...293..424Z}
{Zinn} R., 1985, \apj, 293, 424

\bibitem[{{Zinn} \& {West}(1984)}]{1984ApJS...55...45Z}
{Zinn} R., {West} M.~J., 1984, \apjs, 55, 45

\end{thebibliography}

\appendix
\section{Empirical Colour Transformations}

\label{colours}

For galaxies and GCs that lack $gri$ imaging we used empirical colour conversions to give all GCs equivalent $(g-i)$ colours and $i$ magnitudes.
To convert HST ACS $gz$ photometry into Suprime-Cam $gi$ photometry we derived the following relations using 169 spectroscopically confirmed GCs with both ACS and Suprime-Cam  photometry in NGC 4365: 
\begin{equation}\label{eq:4365cal}
(g - i) = (0.735 \pm 0.009) \times (g - z) + (0.147 \pm 0.012) \text{ mag}
\end{equation}
and:
\begin{equation}\label{eq:4365cali}
i = z + (0.554 \pm 0.032) \times (g - z) + (-0.542 \pm 0.034) \text{ mag .}
\end{equation}
These relations were used to convert the ACS $gz$ photometry in NGC 4278 and NGC 4365.
We also used the same objects to derive a relation between $(g - r)$ and $(g - i)$:
\begin{equation}
(g - i) = (1.639 \pm 0.011) \times (g - r) + (-0.127 \pm 0.007)\text{ mag .}
\end{equation}

To convert $(V-I)$ colours into $(g-i)$ we derived the following relation using 30 photometric GC candidates in NGC 5846 brighter than $i$ = 23:
\begin{equation}\label{eq:5846cal}
(g - i) = (1.259 \pm 0.076) \times (V - I) - (0.411 \pm 0.069)\text{ mag .}
\end{equation}
To convert $I$ to $i$ we used the same data to get:
\begin{equation}
i = I + (0.169 \pm 0.091) \times (V - I) + (0.143 \pm 0.099) \text{ mag .}
\end{equation}
We used these equations to convert the WFPC2 $VI$ photometry in NGC 5846 and NGC 7457.

To convert $BI$ photometry to $gi$ we used the ACS $BI$ photometry of \citet{2006MNRAS.366.1230F} of NGC 1407 and 57 spectroscopically confirmed GCs in NGC 1407 to find: 
\begin{equation}
(g - i) = (0.651 \pm 0.015) \times (B - I) + (-0.187 \pm 0.025) \text{ mag}
\end{equation}
and:
\begin{equation}
i = I + (0.162 \pm 0.146) \times (B - I) + (0.081 \pm 0.259) \text{ mag .}
\end{equation}
We used these relations to convert the ACS $BVI$ photometry in NGC 2768.
Using 30 spectroscopically confirmed GCs in NGC 2768 with both transformed ACS and Suprime-Cam photometry we derived a relation between $(r - z)$ and $(g - i)$:
\begin{equation}
(g - i) = (0.913 \pm 0.227) \times (r - z) + (0.408 \pm 0.154) \text{ mag .}
\end{equation}
We used this relation to convert the colours of GCs with only $Riz$ Suprime-Cam photometry in NGC 2768.

To convert $BVR$ photometry into a $gi$ photometry we used the WIYN Minimosaic $BVR$ photometry of NGC 821 from \citet{2008MNRAS.385..361S} together with 42 spectroscopically confirmed GCs to get the following:
\begin{equation}\label{eq:0821cal}
(g - i) = (1.457 \pm 0.101) \times (B - R) + (-0.878 \pm 0.103) \text{ mag}
\end{equation}
and: 
\begin{equation}
i = R + (-0.565 \pm 0.086) \times (B - R) + (0.385 \pm 0.084)\text{ mag .}
\end{equation}
We used this to convert the WIYN Minimosaic $BVR$ photometry of NGC 7457.

For NGC 4278 \citet{Vince2012} prefer ACS photometry to Suprime-Cam photometry.
When ACS photometry is unavailable they convert the Suprime-Cam $BVI$ photometry into $gz$.
We used Equations \ref{eq:4365cal} and \ref{eq:4365cali} to convert $gz$ into $gi$.

\section{Empirical Colour-Metallicity Relations}
\label{othercolourmetals}
Using the preceding colour--colour relations we converted Equation~\ref{eq:colourmetal} into other optical colours: 
\begin{equation}
\text{[Z/H]} = \left\{\begin{array}{r} (5.47 \pm 0.94) \times (g - z) + (-5.95 \pm 1.02) \\
(g - z) < 0.84 \\
\\
(2.56 \pm 0.09) \times (g - z) + (-3.50 \pm 0.11) \\
(g - z) > 0.84 \end{array}\right.
\end{equation}
\begin{equation}
\text{[Z/H]} = \left\{\begin{array}{r} (12.23 \pm 2.10) \times (g - r) + (-8.04 \pm 1.01) \\
(g - r) < 0.55 \\
\\
(5.72 \pm 0.20) \times (g - r) + (-4.47 \pm 0.11) \\
(g - r) > 0.55 \end{array}\right.
\end{equation}
\begin{equation}
\text{[Z/H]} = \left\{\begin{array}{r} (9.36 \pm 1.71) \times (V - I) + (-0.16 \pm 1.24) \\
(V - I) < 0.94 \\
\\
(4.39 \pm 0.30) \times (V - I) + (-5.46 \pm 0.27) \\
(V - I) > 0.94 \end{array}\right.
\end{equation}
\begin{equation}
\text{[Z/H]} = \left\{\begin{array}{r} (4.86 \pm 0.84) \times (B - I) + (-8.48 \pm 1.04) \\
(B - I) < 1.47 \\
\\
(2.27 \pm 0.09) \times (B - I) + (-4.68 \pm 0.14) \\
(B - I) > 1.47 \end{array}\right.
\end{equation}

\label{lastpage}
\end{document}